\documentclass[journal]{IEEEtran}
\usepackage{graphicx}
\usepackage{color}
\usepackage{placeins}
\usepackage{float}
\usepackage{tabularx,colortbl}
\usepackage{graphics, subfigure, theorem, times, amsfonts, amsmath, amssymb, cite, verbatim}
\usepackage{multicol,multirow,booktabs}
\usepackage{tikz}
\usepackage{framed}
\usetikzlibrary{shapes,arrows}
\tikzstyle{phantom vertex} = [ ellipse, 
                               anchor = center, 
                               minimum height = 0.05*\unit, 
                               minimum width  = 0.05*\unit,
                               inner sep=0pt,
                               anchor=center]
\tikzstyle{red vertex}   = [black, fill = red!20,   phantom vertex, draw]
\tikzstyle{black vertex} = [black, fill = black!20, phantom vertex, draw]
\tikzstyle{blue vertex}  = [black, fill = blue!20,  phantom vertex, draw]
\tikzstyle{green vertex} = [black, fill = green!20,  phantom vertex, draw]
\tikzstyle{yellow vertex} = [black, fill = yellow!20,  phantom vertex, draw]
\tikzstyle{vertex}       = [draw, phantom vertex]

\tikzstyle{point} = [ellipse, inner sep=0pt, draw, anchor = center,
                     minimum height = 0.12*\unit, minimum width  = 0.12*\unit]

\usepackage{pgfplots}
\usepackage{color}
\usepackage{hyperref}
\usepackage{needspace}

\renewcommand{\QED}{\hfill\ensuremath{\blacksquare}}
\newenvironment{myproof}[1][$\!\!$]{{\noindent\bf Proof #1: }}
                         {\hfill\QED\medskip}

\newenvironment{mylist}{\begin{list}{}{  \setlength{\itemsep  }{2pt} \setlength{\parsep    }{0in}
                                         \setlength{\parskip  }{0in} \setlength{\topsep    }{5pt}
                                         \setlength{\partopsep}{0in} \setlength{\leftmargin}{2pt}
                                         \setlength{\labelsep }{5pt} \setlength{\labelwidth}{-5pt}}}
                          {\end{list}\medskip}

\newcounter{excercise}
\newcounter{excercisepart}

\definecolor{pennblue}{cmyk}{1,0.65,0,0.30}
\definecolor{pennred}{cmyk}{0,1,0.65,0.34}
\definecolor{mygreen}{rgb}{0.10,0.50,0.10}
\newcommand \red[1]         {{\color{red}#1}}

\newcommand \blue[1]        {{\color{blue}#1}}

\newcommand \green[1]       {{\color[rgb]{0.10,0.50,0.10}#1}}

\def \reals    {{\mathbb R}}

\def\ccalC{{\ensuremath{\mathcal C}}}
\def\ccalD{{\ensuremath{\mathcal D}}}

\def\ccalH{{\ensuremath{\mathcal H}}}
\def\ccalI{{\ensuremath{\mathcal I}}}

\def\ccalM{{\ensuremath{\mathcal M}}}
\def\ccalN{{\ensuremath{\mathcal N}}}

\def\ccalU{{\ensuremath{\mathcal U}}}

\def\ccal0{{\ensuremath{\mathcal 0}}}
\def\hhatc{{\ensuremath{\hat c}}}
\def\hhatd{{\ensuremath{\hat d}}}

\def\tdC{{\ensuremath{\tilde C}}}

\def\tdZ{{\ensuremath{\tilde Z}}}

\def\tdd{{\ensuremath{\tilde d}}}

\def\tdx{{\ensuremath{\tilde x}}}

\def\bbI{{\ensuremath{\mathbf I}}}

\def\bbp{{\ensuremath{\mathbf p}}}

\def\bb0{{\ensuremath{\mathbf 0}}}
\def\bbepsilon{\boldsymbol{\epsilon}}

\def\bbphi{\boldsymbol{\phi}}

\def \eps    {\epsilon}

\newtheorem{proposition}{\hspace{0pt}\bf Proposition}

\newtheorem{theorem}{\hspace{0pt}\bf Theorem}
\newtheorem{corollary}{\hspace{0pt}\bf Corollary}
\newtheorem{fact}{\hspace{0pt}\bf Fact}
\newtheorem{remark}{\hspace{0pt}\bf Remark}

\newtheorem{definition}{\hspace{0pt}\bf Definition}

\usepackage{accents}

\def \du {\bar {d}}
\def \dl {\underline{d}}

\def \etau {\bar {\eta}}
\def \etal {\underline{\eta}}
\def \cu {\bar {c}}
\def \cl {\underline{c}}

\def \au {\bar {a}}
\def \al {\underline{a}}
\def \bu {\bar {b}}
\def \bl {\underline{b}}
\def \eu {\bar {e}}
\def \el {\underline{e}}
\def \fu {\bar {f}}
\def \fl {\underline{f}}
\def \gu {\bar {g}}
\def \gl {\underline{g}}
\def \hu {\bar {h}}
\def \hl {\underline{h}}

\def \d {\text{d}}
\def \c {\text{c}}

\def \dotx {\dot x}

\def \SL {\textmd{SL}}

\def \sep {\textmd{sep}}

\def \phixx {\phi_{x, x'}}

\def \CO {\textmd{CO}}

\def \CL {\textmd{CL}}

\begin{document}

\title{Hierarchical Clustering Given Confidence \\  Intervals of Metric Distances}

\author{Weiyu Huang and Alejandro Ribeiro
\thanks{Supported by NSF CCF-1217963. Dept. of Electrical and Systems Eng., University of Pennsylvania, 200 S 33rd St., Philadelphia, PA 19104. Email: \{whuang, aribeiro\}@seas.upenn.edu. Preliminary results appeared in \cite{huang16b, huang16c}.}}

\maketitle

\begin{abstract}
This paper considers metric spaces where distances between a pair of nodes are represented by distance intervals. The goal is to study methods for the determination of hierarchical clusters, i.e., a family of nested partitions indexed by a resolution parameter, induced from the given distance intervals of the metric spaces. Our construction of hierarchical clustering methods is based on defining admissible methods to be those methods that abide to the axioms of value -- nodes in a metric space with two nodes are clustered together at the convex combination of the distance bounds between them -- and transformation -- when both distance bounds are reduced, the output may become more clustered but not less. Two admissible methods are constructed and are shown to provide universal upper and lower bounds in the space of admissible methods. Practical implications are explored by clustering moving points via snapshots and by clustering networks representing brain structural connectivity using the lower and upper bounds of the network distance. The proposed clustering methods succeed in identifying underlying clustering structures via the maximum and minimum distances in all snapshots, as well as in differentiating brain connectivity networks of patients from those of healthy controls.
\end{abstract}


\IEEEpeerreviewmaketitle

%
\section{Introduction}\label{sec_introduction}

We often encounter datasets representing points in a metric space but in which the computation of exact distances between points is intractable. When this happens it is customary to resort to tractable lower and upper bounds. This is the case when, e.g., the points themselves represent individual unlabeled networks (Section \ref{sec_real_world}). The space of networks can be endowed with a metric that is computationally intractable because unlabeled networks are invariant to permutations \cite{Huang15}. However, lower bounds are readily available by looking at specific permutations and upper bounds can be computed using homological features \cite{Huang16a}. In this paper we study hierarchical clustering methods for problems of this form. I.e., we want to hierarchically cluster points in a metric space in which the exact distances between pairs of points are not perfectly known but known to belong to some interval.

The approach we take is axiomatic in nature and builds on the increasingly strong theoretical understanding of clustering methods \cite{von2005, ben2006, guyon2009, ben2009, zadeh2009, carlsson2010, carlsson2013, carlsson2014}. Our particular interest here is in hierarchical clustering where instead of a single partition, we search for a family of partitions indexed by a connectivity parameter, e.g., \cite{lance1967, jain1988, zhao2005}. It has been proved in \cite{carlsson2010} that single linkage \cite[Ch. 4]{jain1988} is the unique hierarchical clustering method that abides to three reasonable axioms. These results were later extended to asymmetric networks not necessarily metric and the number of of axioms required for unicity results reduced to only two \cite{carlsson2010, carlsson2014}. In the case of metric spaces the two properties that are imposed as axioms in \cite{carlsson2014} can be intuitively stated as: (A1) The nodes in a network with two nodes are clustered at the resolution specified by their distance. (A2) A network that is uniformly dominated by another should have clusters that are also uniformly dominated. Property (A1) is dubbed the Axiom of Value and property (A2) the Axiom of Transformation. 

The goal of this paper is to extend the axiomatic construction of hierarchical clustering in \cite{carlsson2010, carlsson2014} for clustering based on distance intervals. Condition (A2) is kept with a minor adaptation. To adapt condition (A1) we introduce a confidence parameter which is intended to assign different relative trusts to lower and upper bounds and require that: (A1) The nodes in a network with two nodes are clustered at the convex combination of the interval extremes dictated by the confidence parameter. The contributions of this paper are: (i) To define the {\it combine-and-cluster} and {\it cluster-and-combine} methods that satisfy axioms (A1) and (A2). (ii) To prove that these methods are extremal across all methods that satisfy axioms (A1) and (A2). (iii) To demonstrate the practical applicability of the methods in the clustering of moving points via snapshots and the clustering of networks representing brain structural connectivity. We point out that clustering based on distance intervals is a particular case of the problem of clustering with uncertain observations where the unpredictability is given by the distance intervals. Clustering methods that attempt to take uncertainty into consideration include the construction of models to replicate the properties of uncertainties in the data \cite{cormode2008, gullo2008, schubert2015} as well as the consideration of multiple observations of points given in a Euclidean space \cite{chavent2002, de2004, yu2011, jiang2013}. Our work differs in that we investigate situations where the only available information are the upper and lower bounds of the actual metric distances. This can be considered as a more crude observation and a generalization of the approaches in \cite{cormode2008, gullo2008, schubert2015, chavent2002, de2004, yu2011, jiang2013}.

We begin the paper by visiting necessary definition of hierarchical clustering, dendrograms, ultrametrics, and chains (Section \ref{sec_preliminaries}). We then state formally the axioms of value -- nodes in a metric space with two nodes are clustered at the convex combination of the distance bounds between them -- and transformation -- a network where the lower and upper bounds are uniformly dominated by the respective lower and upper bounds of another network should have clusters that are uniformly dominated by this second network (Section \ref{sec_axioms}). We further demonstrate that the two axioms combined yield another intuitive property that no pairs should be clustered together at a resolution smaller than a given threshold (Section \ref{sec_minimum_separation}). Within this axiomatic framework we construct the {\it combine-and-cluster} and {\it cluster-and-combine} methods (Section \ref{sec_extremal_ultrametrics}). Both of these methods rely on single linkage but differ on the chain cost that is measured to determine if nodes are clustered or not. In combine-and-cluster pairwise distances are estimated by the convex combination of lower and upper bounds and the cost of a chain is the maximum resulting distance. In cluster-and-combine we compute separate chain costs for the lower and upper bounds that are then reduced to their convex combination. We then introduced our main theoretical contribution contribution of the paper by showing that combine-and-cluster and cluster-and-combine provide upper and lower bounds on all methods that are admissible with respect to the axioms of value and transformation (Section \ref{sec_extremal_ultrametrics}). This enables us to characterize the space of admissible methods for metric spaces with distances specified by intervals and draw connections with admissible methods for metric spaces (Section \ref{sec_alpha_0_1}). Practical usefulness of the methods in both synthetic scenarios (Section \ref{sec_synthetic}) as well as real world settings (Section \ref{sec_real_world}) are presented.

%
\section{Preliminaries}\label{sec_preliminaries}

We consider a metric space $M_X$ to be a pair $(X, d_X)$ where $X$ is a finite set of nodes and $d_X: X \times X \rightarrow \reals_+$ is a metric distance. In specific, $d_X(x, x')$ between nodes $x \in X$ and $x' \in X$ is assumed to be nonnegative for all pairs $x, x'$, is symmetric such that $d_X(x, x') = d_X(x', x)$, and is $0$ if and only if the nodes coincide with $x = x'$; $d_X$ also satisfies triangle inequality with $d_X(x, x'') \leq d_X(x, x') + d_X(x', x'')$ for any triplets $x, x', x'' \in X$. The interest of study in this paper is not on the metric space $M_X$, but in scenarios where observation of $d_X(x, x')$ is not exact but given in a confidence interval. Formally, we consider $I_X$ as the triplet $(X, \du _X, \dl _X)$ where $\du_X:X \times X \rightarrow \reals_+$ is an upper bound of the original metric distance and $\dl_X:X \times X \rightarrow \reals_+$ is a lower bound of the metric. Given a pair of nodes $x, x'\in X$, we therefore have the relationship $0 < \dl _X(x, x') \leq d_X(x, x') \leq \du _X(x, x')$. The bounds $\dl(x, x')$ as well as $\du(x, x')$ between nodes $x, x' \in X$ are nonnegative for all pairs and are $0$ if and only if $x = x'$; moreover, they are symmetric, i.e. $\dl(x, x')$ is the same as $\dl(x', x)$ and similarly for $\du(x, x')$.  However, they they may not necessarily satisfy the triangle inequality. We define $\ccalI$ as the set of all metric spaces where the actual distance is observed in a confidence intervals. Entities in $\ccalI$ may have different node sets $X$ as well as different distance lower or upper bounds.

%
\begin{figure}[t]
\centerline{

\def \thisplotscale {1.15}

\def \unit {\thisplotscale cm}

\tikzstyle {vertex} = [circle, 
                       draw,
                       minimum width = 0.7*\unit,
                       minimum height = 0.7*\unit,
                       anchor=center,
                       font=\small]
\tikzstyle {light}  = [opacity = 0.1]
\tikzstyle{bigvertex} = [vertex, 
                        minimum width  = 0.7*\unit,
                        minimum height = 0.7*\unit]

{\footnotesize \begin{tikzpicture}[x = 1.5*\unit, y = 0.7*\unit]

	\node at (9.8, 0) (center1) {};
	\path (center1) ++ ( 135:1.401) node (1) [fill = blue!20, vertex] {$a$};
	\path (center1) ++ (  45:1.401) node (2) [fill = blue!20, vertex] {$b$};
	\path (center1) ++ (- 45:1.401) node (3) [fill = blue!20, vertex] {$c$};
	\path (center1) ++ (-135:1.401) node (4) [fill = blue!20, vertex] {$d$};
	\path[-stealth, -] (1) edge [bend left,  above] node {$\du = 7$} (2);
	\path[-stealth, -] (1) edge [bend left,  line width = 0pt, below] node {$\dl = 1$} (2);
	\path[-stealth, -] (2) edge [bend left,  right] node {$\du = 5$} (3);
	\path[-stealth, -] (2) edge [bend left,  line width = 0pt, left] node {$\dl = 3$} (3);
	\path[-stealth, -] (3) edge [bend left,  below] node {$\du = 4$} (4);
	\path[-stealth, -] (3) edge [bend left,  line width = 0pt, above] node {$\dl = 2$} (4);
	\path[-stealth, -] (4) edge [bend left,  left ] node {$\du = 6$} (1);
	\path[-stealth, -] (4) edge [bend left,  line width = 0pt, right] node {$\dl = 3$} (1);
	\path[-stealth, -] (1) edge [gray, above] node {$\du = 8$} (3);
	\path[-stealth, -] (1) edge [gray, line width = 0pt, below] node {$\dl = 8$} (3);
	\path[-stealth, -] (2) edge [gray, above] node {} (4);
	\path[-stealth, -] (2) edge [gray, line width = 0pt, below] node {} (4);

\end{tikzpicture}}  
}
\caption{An example of metric space where distances between pairs of nodes are given in lower and upper bounds. The intuition of clustering is ambiguous because the notion of proximity is no longer clear, e.g. the pair $a, b$ has the smallest distance lower bound whereas the pair $c,d$ has the smallest average of their distance lower and upper bounds. It is not clear which of the two pairs is more proximate.
}
\label{fig_interval_example}
\end{figure}
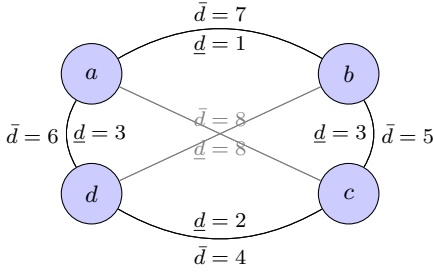

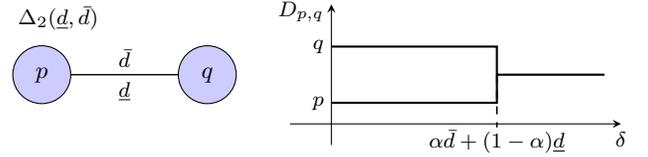
\begin{figure}[t]
\centerline{

\def \thisplotscale {1.1}

\def \unit {\thisplotscale cm}

\tikzstyle {vertex} = [circle, 
                       draw,
                       minimum width = 0.7*\unit,
                       minimum height = 0.7*\unit,
                       anchor=center,
                       font=\small]
\tikzstyle {light}  = [opacity = 0.1]
\tikzstyle{bigvertex} = [vertex, 
                        minimum width  = 0.7*\unit,
                        minimum height = 0.7*\unit]

{\footnotesize \begin{tikzpicture}[x = 1*\unit, y = 0.85*\unit]

	\node at (0, 0) (center1) {};
	\path (center1) ++ ( 180:1) node (1) [fill = blue!20, vertex] {$p$};
	\path (center1) ++ ( 0:1) node (2) [fill = blue!20, vertex] {$q$};
	\path (1) ++ (0.2 , 0.8) node (title) [] {$\Delta_2(\dl, \du)$};
	\path[-stealth, -] (1) edge [above] node {$\du$} (2);
	\path[-stealth, -] (1) edge [line width = 0pt, below] node {$\dl$} (2);

   
    \draw [-stealth] (2,-0.7) -- (6,-0.7) node [below, at end] {$\delta$};
    \draw [-stealth] (2.5,-1) -- (2.5,1);
    
    \draw[thick, -] (2.5,-0.4) -- ++(2,0) -- ++(0,0.8) -- ++(-2,0) ++(2, -0.4) --++(1.3, 0);
    \node [left] at (2.5, -0.4) {$p$};
    \node [left] at (2.5,0.4) {$q$};
    \node [left] at (2.5,0.9) {$D_{p, q}$};
    
    \draw [-stealth, dashed, -, semithick] (4.5,-0.0) -- (4.5,-0.75);
    \node [] at (4.5,-0.95) {$\alpha\du + (1 - \alpha) \dl$};

\end{tikzpicture}} }
\caption{Two-node space $\Delta_2(\du, \dl)$ and the Axiom of Value: nodes in a two-node space cluster at the convex combination of the distance upper and lower bounds.}
\label{fig_two_node_space}
\end{figure}

%
An example metric space with distance given by intervals is shown in Fig.~\ref{fig_interval_example}. The set of nodes is $X = \{a, b, c, d\}$ with distance upper and lower bounds represented by values adjacent to each edge. The lower bound $\dl_X(a, b)$ of distance, e.g. from $a$ to $b$ is $1$, is smaller than the distance upper bound $\du_X(a,b) = 7$. The smallest nontrivial case contains two nodes $p$ and $q$ with distance lower bound $\dl(p, q) = \dl$ as well as upper bound $\du(p, q) = \du \geq \dl > 0$ is described in Fig.~\ref{fig_two_node_space}. This special space appears often in the proceeding discussion of the paper, and we define the two-node space $\Delta_2(\dl, \du)$ with parameters $\dl$ and $\du$ as
\begin{align}\label{eqn_two_node_space} 
    \Delta_2(\dl, \du) := (\{p, q\}, \dl, \du).
\end{align}

A clustering of the set $X$ denotes a partition $P_X$ of $X$, i.e. a collection of pairwise disjoint sets $P_X = \{S_1, S_2, \dots, S_J\}$ with $S_i \cap S_j = \emptyset$ for any $i \neq j$ are required to cover $X$, $\cup_{j = 1}^J S_j = X$. The sets $S_1, \dots, S_J$ are named the clusters of $P_X$. An equivalence relation $\sim$ on $X$ is a binary relation such that for all $x, x', x'' \in X$ we have that (1) $x \sim x$, (2) $x \sim x'$ if and only if $x' \sim x$, and (3) $x \sim x'$ and $x' \sim x''$ would imply $x \sim x''$. A partition $P_X = \{S_1, S_2, \dots, S_J\}$ of $X$ always induces and is induced by an equivalence relation $\sim_{P_X}$ on $X$ where for all $x, x' \in X$ we have that $x \sim_{P_X} x'$ if and only if $x$ and $x'$ is cluttered to the same set $S_j$ for some $j$. 

In this paper we focus on hierarchical clustering methods \cite{carlsson2010, carlsson2014}. The output of hierarchical clustering methods is not a single partition $P_X$ but a nested collection $\ccalD_X$ of partitions $D_X(\delta)$ of $X$ indexed by the resolution parameter $\delta \geq 0$. In the language of equivalence relation defined previously, for a given $\ccalD_X$, we say that two nodes $x$ and $x'$ are equivalent at resolution $\delta$ with notation $x \sim_{D_X(\delta)} x'$ if and only if nodes $x$ and $x'$ are in the same cluster of $D_X(\delta)$. The nested collection $\ccalD_X$ is named a {\it dendrogram} and is required to satisfy the following property \cite{carlsson2010}:

\begin{mylist}
\item [{\it (D1) Boundary conditions}.] For $\delta = 0$ the partition $D_X(0)$ clusters each $x \in X$ into a separate singleton and for some $\delta_\infty$ sufficiently large $D_X(\delta_\infty)$ clusters all elements into a single set.
\item [{\it (D2) Hierarchy}.] As $\delta$ increases clusters can be combined but not separated. I.e., for any $\delta < \delta'$, any given pair of nodes $x, x'$ with $x \sim_{D_X(\delta)} x'$ would satisfy $x \sim_{D_X(\delta')} x'$.
\item [{\it (D3) Right continuity}.] For all $\delta \geq 0$, there exists an $\epsilon > 0$ such that $D_X(\delta') = D_X(\delta) $ for any $\delta' \in [\delta, \delta + \eps]$.
\end{mylist}

The interpretation of a dendrogram is that of a structure which yields different clustering results at different resolutions. At resolution $\delta = 0$ each node is in a cluster of its own. As the resolution parameter $\delta$ increases, nodes start forming clusters. Based on the condition (D2), nodes become more clustered since once they join together in a cluster at some resolution, they stay together in the same cluster for all larger resolutions. Denote $\ccalD$ as the space of all dendrograms, hierarchical clustering method upon distance intervals is defined as a function $\ccalH: \ccalI \rightarrow \ccalD$ from the space $\ccalI$ to the space of dendrograms $\ccalD$ such that the underlying space $X$ is preserved. For the triplet $I_X = (X, \dl_X, \du_X)$, we denote $D_X = \ccalH(X, \dl_X, \du_X)$ as the output of $\ccalH$.

%

%
\subsection{Dendrograms as Ultrametrics}\label{sec_ultrametrics}

Dendrograms are difficult to analyze. A more convenient representation is acquired when dendrograms are identified with finite ultrametric spaces. An ultrametric on the space $X$ is a metric $u_X: X \times X \rightarrow \reals_+$ satisfying the stronger triangle inequality such that any points $x, x', x'' \in X$, the ultrametrics $u_X(x, x'')$, $u_X(x, x')$, and $u_X(x', x'')$ abide to the relationship
\begin{align}\label{eqn_ultrametrics}
    u_X(x, x'') \leq \max \big( u_X(x, x'), u_X(x', x'')\big).
\end{align}
Ultrametric spaces are particular cases of metric spaces since \eqref{eqn_ultrametrics} would imply the usual triangle inequality $u_X(x, x'') \leq u_X(x, x') + u_X(x', x'')$. We investigate ultrametrics because a structure preserving bijective mapping between dendrograms and ultrametrics can be established \cite{carlsson2010}. Consider the map $\Phi: \ccalD \rightarrow \ccalU$ from the space of dendrograms to the space of ultrametrics: given a dendrogram $\ccalD_X$ over a finite set $X$, the output $\Phi(D_X) = (X, u_X)$ with $u_X(x,x')$ for any pair of nodes $x, x' \in X$ is defined as the smallest resolution at which $x$ and $x'$ are clustered together 
\begin{align}\label{eqn_map_dendrogram_ultrametric}
    u_X(x, x') := \min \left\{ \delta > 0 : x \sim_{D_X(\delta)} x' \right\}.
\end{align}
The map $\Psi: \ccalU \rightarrow \ccalD$ is constructed such that for a given ultrametric space $(X, u_X)$ and any resolution $\delta \geq 0$, the equivalence relationship $\sim_{u_X(\delta)}$ is defined as
\begin{align}\label{eqn_map_ultrametric_dendrogram}
    x \sim_{u_X(\delta)} x' \Leftrightarrow u_X(x, x') \leq \delta.
\end{align}
Denote the cluster result at $\delta$ as $D_X(\delta) := \{ X \mod \sim_{u_X(\delta)} \}$ where nodes belonging to the same equivalence class is clustered together. The output of the map is then $\Psi(X, u_X) := \ccalD_X$. It is shown \cite{carlsson2010} that the maps defined above preserve structures in the respective space as we state in the following theorem.

%
\begin{theorem}\label{thm_deodemgram_ultrametric}
The maps $\Phi: \ccalD \rightarrow \ccalU$ and $\Psi:\ccalU \rightarrow \ccalD$ are both well defined. Moreover, $\Phi \circ \Psi$ is the identity on $\ccalU$ and $\Psi \circ \Phi$ is the identity on $\ccalD$.
\end{theorem}

%
Given the equivalence between dendrograms and ultrametrics demonstrated by Theorem \ref{thm_deodemgram_ultrametric} we can consider hierarchical clustering methods $\ccalH$ as inducing ultrametrics in node spaces $X$ based on distance intervals $\dl_X$ and $\du_X$ and reinterpret the method $\ccalH$ as a map $\ccalH: \ccalI \rightarrow \ccalU$ from the space of metric spaces given confidence intervals to the space of ultrametrics. The outcome of a hierarchical clustering method constructs an ultrametric in the same space $X$ even when the original observation is given as distance intervals of a metric distance. We say that two clustering methods $\ccalH_1$ and $\ccalH_2$ being equivalent with notation $\ccalH_1 \equiv \ccalH_2$ if and only if $\ccalH_1(I) = \ccalH_2(I)$ for any $I \in \ccalI$.

%
\begin{figure}[t]
{\center

\def \thisplotscale {0.78}

\def \unit {\thisplotscale cm}

\tikzstyle {vertex} = [circle, 
                       draw,
                       minimum width = 0.7*\unit,
                       minimum height = 0.7*\unit,
                       anchor=center,
                       font=\small]
\tikzstyle {light}  = [opacity = 0.1]
\tikzstyle{bigvertex} = [vertex, 
                        minimum width  = 0.7*\unit,
                        minimum height = 0.7*\unit]

{\scriptsize \begin{tikzpicture}[x = 0.65*\unit, y = 0.5*\unit]

    
    \node at (1.5, 2) (center1) {};
    \path (center1) ++ ( 90:2.5) node (1) [fill = blue!20, vertex] {$x_1$};
    \path (center1) ++ ( -45:2.5) node (2) [fill = blue!20, vertex] {$x_2$};
    \path (center1) ++ ( -135:2.5) node (3) [fill = blue!20, vertex] {$x_3$};


    \node at (8, 2) (center1) {};
    \path (center1) ++ ( 90:1.6) node (1p) [fill = blue!20, vertex] {$x_1$};
    \path (center1) ++ ( -45:1.6) node (2p) [fill = blue!20, vertex] {$x_2$};
    \path (center1) ++ ( -135:1.6) node (3p) [fill = blue!20, vertex] {$x_3$};
    
    \path (1) edge [-stealth, bend left=20, above, red, very thick, pos=0.6] node {$\bbphi$} (1p);	
    \path (2) edge [-stealth, bend right=35, above, red, very thick, pos=0.5] node {$\bbphi$} (2p);	    
    \path (3) edge [-stealth, bend left=12, above, red, very thick, pos=0.75] node {$\bbphi$} (3p);	   

    \path[-stealth, -] (1) edge [bend left,  right, pos = 0.3] node {$\du = 2$} (2);
    \path[-stealth, -, line width = 0pt] (1) edge [bend left,  right, pos = 0.7] node {$\dl = 1$} (2);
    \path[-stealth, -] (2) edge [bend left,  below] node {$\du = 3$} (3);
    \path[-stealth, -, line width = 0pt] (2) edge [bend left,  above] node {$\dl = 2$} (3);
    \path[-stealth, -] (3) edge [bend left,  left, pos = 0.7] node {$\du = 3$} (1);
    \path[-stealth, -, line width = 0pt] (3) edge [bend left,  left, pos = 0.3] node {$\dl = 2$} (1);

    \path[-stealth, -] (1p) edge [bend left,  right, pos = 0.2] node {$\du = 1$} (2p);
    \path[-stealth, -, line width = 0pt] (1p) edge [bend left,  right, pos = 0.8] node {$\dl = 1/2$} (2p);
    \path[-stealth, -] (2p) edge [bend left,  below] node {$\du = 1$} (3p);
    \path (center1) ++ ( -90:3) node {$\dl = 1/2$};
    \path[-stealth, -] (3p) edge [bend left, left, pos = 0.8] node {$\du = 1$} (1p);
    \path[-stealth, -, line width = 0pt] (3p) edge [bend left, left, pos = 0.2] node {$\dl = 1/2$} (1p);

   
    \draw [-stealth] (12.5,2) -- (16.3,2) node [below, at end] {$\delta$};
    \draw [-stealth] (13,2) -- (13,4.9);
    
    \draw[thick, -] (13,2.4) -- ++(1.5,0) -- ++(0,0.8) -- ++(-1.5,0) ++(0,0.8) -- ++(2.5,0) -- ++(0,-1.2) -- ++(-1,0) -- ++(1,0)--++ (0, 0.6)--++(1, 0);
    \node [left] at (13,2.4) {$x_1$};
    \node [left] at (13,3.2) {$x_2$};
    \node [left] at (13,4) {$x_3$};
    \node [left] at (12.3,4.5) {$D_{X}$};
    
    \node at (8.5,4.7) {$N_{Y}$};
     \node at (0.5,4.7) {$N_{X}$};
    
    
    \draw [-stealth] (12.5,-1) -- (16.3,-1) node [below, at end] {$\delta$};
    \draw [-stealth] (13,-1) -- (13,1.9);
    
    \draw[thick, -] (13,-0.6) -- ++(1.2,0) -- ++(0,0.8) -- ++(-1.2,0) ++(0,0.8) -- ++(1.2,0) -- ++(0,-0.8) -- ++(2.1,0);
    \draw[dashed, thick, red](15,4.5) -- ++(0,-6) node [right, at end] {$\delta'$};
    \node [left] at (13,-0.6) {$y_1$};
    \node [left] at (13,0.2) {$y_2$};
    \node [left] at (13,1) {$y_3$};
    \node [left] at (12.3,1.5) {$D_{Y}$};

\end{tikzpicture}
}}
\caption{Axiom of Transformation. If $I_X$ can be mapped to $I_Y$ using a $\alpha$-distance-reducing map $\phi$ [cf. \eqref{eqn_distance_reducing_d} and \eqref{eqn_distance_reducing_c}], then for every resolution $\delta$ nodes clustered together in $D_X(\delta)$ must also be clustered in $D_Y(\delta)$.}
\label{fig_axiom_transformation}
\end{figure}
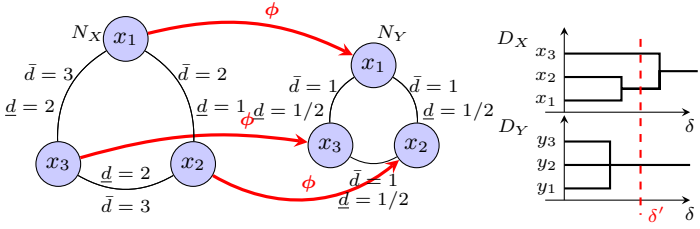

%
\subsection{Chain, Upper and Lower Chain Costs}\label{sec_chains}

The notions of chain and chain cost are substantial in hierarchical clustering. Given a metric space with distance intervals $(X, \dl, \du)$ and a pair of nodes $x, x' \in X$, a {\it chain} from $x$ to $x'$ is any ordered sequence of nodes in $X$,
\begin{align}\label{eqn_chain}
    [x = x_0, x_1, \dots, x_{l-1}, x_l = x'],
\end{align}
which begins with $x$ and ends at $x'$. We denote $C(x, x')$ as one such chain and say $C(x, x')$ connects $x$ to $x'$. Given two chains $C(x, x')$ and $C(x', x'')$ such that the end point $x'$ of the first chain is the same as the starting point of the second, we define the {\it concatenated chain} $C(x, x') \uplus C(x', x)$ as
\begin{align}\label{eqn_concatenated_chain}
    [x = x_0, x_1, \dots, x_{l-1}, x_l = x' = x_0', x_1', \dots, x_{l'}' = x''].
\end{align}
It follows from \eqref{eqn_concatenated_chain} that the concatenation operation $\uplus$ is associative such that $\big(C(x, x') \uplus C(x', x'') \big) \uplus C(x'', x''') = C(x, x') \uplus \big( C(x', x'') \uplus C(x'', x''')  \big)$. For the chain $C(x, x')$, we define its {\it upper cost} and {\it lower cost} respectively as
\begin{align}\label{eqn_upper_lower_costs}
    \max_{i\mid x_i \in C(x, x')} \du_X (x_i, x_{i+1}), \quad 
          \max_{i\mid x_i \in C(x, x')} \dl_X (x_i, x_{i+1}),
\end{align}
i.e. the maximum distance upper or lower bounds encountered as traversing the chain in order. The {\it minimum upper chain cost} $\cu(x, x')$ and the {\it minimum lower chain cost} $\cl(x, x')$ between $x$ and $x'$ is then defined respectively as the minimum upper and lower cost among all chains connecting $x$ to $x'$,
\begin{align}\label{eqn_min_upper_costs}
    \cu_X(x, x') := \min_{C(x, x')} \max_{i\mid x_i \in C(x, x')} \du_X (x_i, x_{i+1}), \\ 
    \cl_X(x, x') := \min_{C(x, x')} \max_{i\mid x_i \in C(x, x')} \dl_X (x_i, x_{i+1}).
    \label{eqn_min_lower_costs}
\end{align}

The minimum upper chain cost $\cu_X(x, x')$ and lower chain cost $\cl_X(x, x')$ are different in general, however they are equal in the degenerate case where distance lower bounds and upper bounds coincide with $\dl_X(x, x') = \du_X(x, x') := d_X(x, x')$ for any $x, x' \in X$. In this case, the minimum cost $\cu_X(x, x') = \cl_X(x, x')$ are important in the construction of the single linkage \cite{carlsson2010}. In specific, single linkage ultrametric $u_X^\SL(x, x')$ between $x$ and $x'$ is 
\begin{equation}\begin{aligned}\label{eqn_single_linkage_ultrametric}
    u_X^\SL(x, x') &= \cu_X(x, x') = \cl_X(x, x') \\
        & = \min_{C(x, x')} \max_{i\mid x_i \in C(x, x')} d_X (x_i, x_{i+1}).
\end{aligned}\end{equation}
In terms of single linkage dendrogram $\SL_X$, for a given resolution $\delta$, the equivalence classes at resolution $\delta$ is
\begin{align}\label{eqn_single_linkage_dendrogram}
    x \sim_{\SL_X(\delta)} x' \Leftrightarrow \cu_X(x, x') = \cl_X(x, x') \leq \delta.
\end{align}
It can be seen that $\cu_X$ is the result of applying single linkage towards the node set $X$ equipped with dissimilarity $\du_X$ despite the fact that $\du_X$ may not be a valid metric; similar result holds for $\cl_X$. In the degenerative case where distance lower bounds and upper bounds coincide, it is equivalent to consider metric spaces $(X, d_X)$. It has been shown \cite{carlsson2010} that single linkage is the unique hierarchical clustering method fulfilling axioms (A1) and (A2) discussed in Section \ref{sec_axioms} plus a third axiom stating that the clusters cannot be formed at resolutions smaller than the minimum distance in the space. In the case when the dissimilarity $d_X(x,x')$ are only given in an interval $[\dl_X(x, x'), \du_X(x,x')]$, the space of methods satisfying axioms (A1) and (A2) and their analogous ones becomes richer, as we explain throughout the paper.

%
\section{Axioms of Value and Transformation}\label{sec_axioms}

To study hierarchical clustering methods on metric spaces where observations of dissimilarities between pairs are given in a distance intervals, we translate natural concepts into the axioms of value and transformation, described in this section. We say a hierarchical clustering method $\ccalH$ is {\it admissible} if and only if it satisfies both the the axioms of transformation and value.

The Axiom of Value is achieved by considering the two-node space $\Delta_2(\dl, \du)$ defined in \eqref{eqn_two_node_space} and described in Fig.~\ref{fig_two_node_space}. In the degenerate special case where $\dl = \du := d(p, q)$, it is apparent that the resolution at which nodes $p$ and $q$ are first clustered together should be $d(p, q)$. In general scenarios where the dissimilarity $d(p, q)$ is given in an interval $[\dl, \du]$ with $\dl < \du$, it is reasonable to consider different resolutions at which nodes $p$ and $q$ start to be in the same cluster. In specific, we say that nodes $p$ and $q$ form a single cluster first at resolution $\delta := \alpha \du + (1 - \alpha) \dl$, the convex combination of the upper and lower bounds $\du$ and $\dl$. Property of hierarchical clustering then indicates nodes $p$ and $q$ are clustered together at any resolution $\delta \geq \alpha \du + (1 - \alpha) \dl$. 
The parameter $\alpha$ controls the level of confidence in examining the distance intervals. A higher value of $\alpha$ implies a more conservative consideration, where in the extreme case with $\alpha = 1$, nodes $p$ and $q$ are clustered together at the distance upper bound $\du$; a lower value of $\alpha$ suggests a more aggressive examination, and in the other extremal scenario with $\alpha = 0$, nodes $p$ and $q$ considered to be in the same cluster as long as the resolution is no smaller than their distance lower bound $\dl$. Since a hierarchical clustering method is a map $\ccalH$ from metric distance intervals to dendrograms, we formalize this intuition as the following requirement.

\begin{mylist}
\item [\it (A1) Axiom of Value.] Given a value $0\leq \alpha \leq 1$, the dendrogram $D_{p, q} = \ccalH(\Delta_2(\dl, \du))$ produced by applying $\ccalH$ to the two-node space $\Delta_2(\dl, \du)$ is such that $D_{p, q}(\delta) = \big\{ \{p\}, \{q\}\big\}$ for $0 \leq \delta < \alpha \du + (1 - \alpha) \dl$ and $D_{p, q}(\delta) = \big\{ \{p, q\} \big\}$ otherwise.
\end{mylist}

One may argue that clustering nodes $p$ and $q$ at any monotone increasing function of $\alpha \du + (1 - \alpha) \dl$ would be admissible. Nonetheless, the current formulation implies that the clustering resolution parameter $\delta$ is expressed in the same units as the distance intervals. From Theorem \ref{thm_deodemgram_ultrametric}, we can rewrite the Axiom of Value by referring to properties of the output ultrametrics.

\begin{mylist}
\item [\it (A1) Axiom of Value.] Given a value $0\leq \alpha \leq 1$, the ultrametric output $(\{p, q\}, u_{p,q}) = \ccalH(\Delta_2(\dl, \du))$ resulted from applying $\ccalH$ upon the two-node space $\Delta_2(\dl, \du)$ satisfies that
\begin{align}\label{eqn_axiom_of_value_ultrametric}
    u_{p, q} (p, q) = \alpha \du + (1 - \alpha) \dl.
\end{align}\end{mylist}

The second requirement on the space of desired methods $\ccalH$ formalizes the intuition for the behavior of $\ccalH$ when considering a transformation w.r.t. the distance upper and lower bounds on the underlying space $X$; see Fig.~\ref{fig_axiom_transformation}. Consider two metric spaces with observations given by distance intervals $I_X = (X, \dl_X, \du_X)$ and $I_Y = (Y, \dl_Y, \du_Y)$ and denote $D_X = \ccalH(X, \dl_X, \du_X)$ and $D_Y = \ccalH(Y, \dl_Y, \du_Y)$ as the corresponding dendrogram outputs. If we can map all the nodes of the triplet $(X, \dl_X, \du_X)$ into nodes of $(Y, \dl_Y, \du_Y)$ such that the combination of lower and upper bounds for any pair of nodes is not increased, we expect the latter metric distance intervals to be more clustered than the former one at any given resolution. Intuitively, nodes in $I_Y$ are less dissimilar with respect to each other, and therefore at any resolution $\delta$ in the respective dendrograms, we expect that for nodes that are clustered in $I_X$, their corresponding nodes in $Y$ are also clustered in $I_Y$. In order to formalize this intuition, we introduce the following notion that given two metric spaces with observations given by distance intervals $I_X = (X, \dl_X, \du_X)$ and $I_Y = (Y, \dl_Y, \du_Y)$ and a value $0 \leq \alpha \leq 1$, the map $\phi: X \rightarrow Y$ is called {\it $\alpha$-distance-reducing} if for any $x, x' \in X$, it holds that
\begin{align}\label{eqn_distance_reducing_d}
    \nonumber & \alpha \du_X(x, x') +  (1 - \alpha) \dl_X(x, x') \\
        &~~~~\geq \alpha \du_Y(\phi(x), \phi(x')) (1 - \alpha) \dl_Y(\phi(x), \phi(x')); \\
            \label{eqn_distance_reducing_c}
                \nonumber & \alpha \cu_X(x, x') +  (1 - \alpha) \cl_X(x, x') \\
                    &~~~~\geq \alpha \cu_Y(\phi(x), \phi(x')) (1 - \alpha) \cl_Y(\phi(x), \phi(x')).
\end{align}
A mapping is $\alpha$-distance-reducing if both the combinations of distance bounds and chain costs is non-increasing. Notice that, in the degenerate case where distance lower and upper bounds coincide, $u_X^\SL(x, x') := \cu_X(x, x') = \cl_X(x, x')$ is the output of applying single linkage upon the metric space. Therefore \eqref{eqn_distance_reducing_d} becomes identical as the requirement $d_X(x, x') \geq d_Y(\phi(x), \phi(x'))$, from which the condition in \eqref{eqn_distance_reducing_c} that $c_X(x, x') \geq c_Y(\phi(x), \phi(x'))$ follows directly. In general cases where distance bounds do not coincide, \eqref{eqn_distance_reducing_c} does not follow from \eqref{eqn_distance_reducing_d} and therefore we need to state both of them. The Axiom of Transformation introduced next is a formal statement of the intinction above.

%
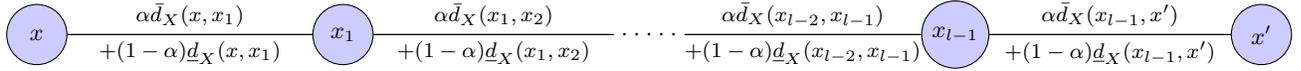
\begin{figure*}[t]
\centerline{

\def \thisplotscale {1}

\def \unit {\thisplotscale cm}

\tikzstyle {vertex} = [circle, 
                       draw,
                       minimum width = 0.8*\unit,
                       minimum height = 0.8*\unit,
                       anchor=center,
                       font=\footnotesize]
\tikzstyle {light}  = [opacity = 0.1]
\tikzstyle{bigvertex} = [vertex, 
                        minimum width  = 0.7*\unit,
                        minimum height = 0.7*\unit]

\def \dis {3.7}
{\footnotesize \begin{tikzpicture}[x = 1.1*\unit, y = 1*\unit]

	\node at (0, 0) (center1) {};
	\path (center1) node (1) [fill = blue!20, vertex] {\small $x$};
	\path (center1) ++ ( \dis, 0) node (2) [fill = blue!20, vertex] {\small $x_1$};
	\path (center1) ++ ( 2*\dis, 0) node (3) {\small $\cdot \cdot \cdot \cdot \cdot$};
	\path (center1) ++ ( 3*\dis, 0) node (4) [fill = blue!20, vertex] {\small $x_{l-1}$};
	\path (center1) ++ ( 4*\dis, 0) node (5) [fill = blue!20, vertex] {\small $x'$};
	\path[-stealth, -] (1) edge [above] node {$\alpha \du_X(x, x_1)$} (2);
	\path[-stealth, line width = 0pt, -] (1) edge [below] node {$+ (1 - \alpha) \dl_X(x, x_1)$} (2);
	\path[-stealth, -] (2) edge [above] node {$\alpha \du_X(x_1, x_2)$} (3);
	\path[-stealth, line width = 0pt, -] (2) edge [below] node {$+ (1 - \alpha) \dl_X(x_1, x_2)$} (3);
	\path[-stealth, -] (3) edge [above] node {$\alpha \du_X(x_{l-2}, x_{l-1})$} (4);
	\path[-stealth, line width = 0pt, -] (3) edge [below] node {$+ (1 - \alpha) \dl_X(x_{l-2}, x_{l-1})$} (4);
	\path[-stealth, -] (4) edge [above] node {$\alpha \du_X(x_{l-1}, x')$} (5);
	\path[-stealth, line width = 0pt, -] (4) edge [below] node {$+ (1 - \alpha) \dl_X(x_{l-1}, x')$} (5);
\end{tikzpicture}} }
\caption{Combine-and-cluster clustering. Nodes $x$ and $x'$ are clustered together at resolution $\delta$ if there exists a chain such that the maximum convex combination of distance bounds $\hhatd_X(x_i, x_{i+1}) = \alpha \du_X(x_i, x_{i+1}) + (1 - \alpha) \dl_X(x_i, x_{i+1})$ is no greater than $\delta$ [cf. \eqref{eqn_combine_and_cluster}]. Of all methods that satisfy the Axioms of Value and Transformation, combine-and-cluster clustering yields the largest ultrametric between any pair of nodes.}
\label{fig_combine_and_cluster}
\end{figure*}

%
\begin{mylist}
\item [\it (A2) Axiom of Transformation.] Consider $I_X = (X, \dl_X, \du_X)$ and $I_Y = (Y, \dl_Y, \du_Y)$ and a $\alpha$-dissimilarity-reducing map $\phi: X \rightarrow Y$. The method $\ccalH$ abides to the axiom of transformation if the dendrograms $D_X = \ccalH(X, \dl_X, \du_X)$ and $D_Y = \ccalH(Y, \dl_Y, \du_Y)$ satisfy for any $\delta \geq 0$, $x \sim_{D_X(\delta)} x'$ implies $\phi(x) \sim_{D_Y(\delta)} \phi(x')$.
\end{mylist}

Rewrite the Axiom of Transformation as in the properties of the output ultrametrics yields the following statement.

\begin{mylist}
\item [\it (A2) Axiom of Transformation.] Consider $I_X = (X, \dl_X, \du_X)$ and $I_Y = (Y, \dl_Y, \du_Y)$ and a given $\alpha$-distance-reducing map $\phi: X \rightarrow Y$. For any pair of nodes $x,x' \in X$, the output ultrametrics $u_X = \ccalH(X, \dl_X, \du_X)$ and $u_Y = \ccalH(Y, \dl_Y, \du_Y)$ satisfy
\begin{align}\label{eqn_axiom_of_transformation_ultrametric}
    u_X(x, x') \geq u_Y(\phi(x), \phi(x')).
\end{align}\end{mylist}

In summary, Axiom (A1) states that the units of the resolution parameter $\delta$ are the same as that of the distance intervals and specifics our tendency in believing lower or upper bounds. Axiom (A2) states that if we reduce both the distance lower and upper bounds, clusters may be combined but cannot be separated. These axioms are an adaption of the axioms proposed in \cite{carlsson2010, carlsson2013} for the degenerate case of $\dl_X = \du_X$ which is equivalent to finite metric spaces, and the axioms proposed in \cite{carlsson2014} for asymmetric networks. 

%
\subsection{Minimum Separation}\label{sec_minimum_separation}

In this subsection we build another intuition on clustering. In the degenerate case where distance lower and upper bounds coincide, it is intuitive that no clusters should be formed at resolutions smaller than the smallest dissimilarity in the metric space. To formalize such intuitive idea, defining {separation} of a given metric space $(X, d_X)$ as the minimum positive distance, 
\begin{align}\label{eqn_separation}
    \sep(X, d_X) := \min_{x \neq x'} d_X(x, x'),
\end{align}
the ultrametrics resulting from any reasonable hierarchical clustering then need to satisfy $u_X(x, x') \geq \sep(X, d_X)$ for any pair of nodes $x \neq x' \in X$. This requirement is stated as an axiom in consideration of clustering methods for metric spaces in \cite{carlsson2010, carlsson2013}. The separation can also be represented in terms of chain costs
\begin{align}\label{eqn_separation_chain}
    \sep(X, d_X) := \min_{x \neq x'} \min_{C(x, x')} \max_{i \mid x_i \in C(x, x')} d_X(x_i, x_{i+1}).
\end{align}
Eq. \eqref{eqn_separation} and \eqref{eqn_separation_chain} are equivalent because for the optimal pair of nodes $\dot x$ and $\dot x'$, the optimal chain $C^\star(\dotx, \dotx')$ would just be the connection $[\dotx, \dotx']$ between them. However, they are different when distance are given in an interval. For general scenarios where the distance upper and lower bounds differ, the $\alpha$ investigated in the Axiom of Value states when nodes in a two-node space should be clustered together. It provides a way to combine the distance bounds and represents where our belief in the distance interval. We would expect a measure defined using $\alpha$ would carry an analogous notion of separation in metric spaces. To do that, we define $\alpha$-separation $s^\alpha_X(x, x')$ between two different nodes $x, x' \in X$ in a metric space with distances given by intervals $(X, \dl_X, \du_X)$ as
\begin{align}\label{eqn_alpha_separation_nodes}
    s^\alpha_X(x, x') = \alpha \cu_X(x, x') + (1 - \alpha) \cl_X(x, x'). 
\end{align}
In words, we search for the optimal chain $C(x, x')$ minimizing the upper chain cost, look for the optimal chain $C'(x, x')$ minimizing the lower chain cost, and take the convex combination of these chain costs. The {\it $\alpha$-separation} for $(X, \dl_X, \du_X)$ is then defined as the minimum of $s^\alpha_X(x, x')$ for all nodes $x \neq x'$
\begin{align}\label{eqn_alpha_separation}
    \sep^\alpha(X, \dl_X, \du_X) := \min_{x \neq x'} s^\alpha_X(x, x').
\end{align}
In the degenerate case we would have $\sep^\alpha(X, \dl_X, \du_X) = \sep(X, d_X)$ for any $\alpha$. Following the notion of separation, for resolutions $0 \leq \delta < \sep^\alpha(X, \dl_X, \du_X)$, no nodes should be clustered together. In the language of ultrametrics, this implies that we must have $u_X(x, x') \geq \sep^\alpha(X, \dl_X, \du_X)$ for any pair of different nodes $x \neq x' \in X$ as we state in the next property.

\begin{mylist}
\item [\it (P1) Property of Minimum Separation.] For $(X, \dl_X, \du_X)$, the output ultrametric $(X, u_X) = \ccalH(X, \dl_X, \du_X)$ of the hierarchical clustering method $\ccalH$ needs to satisfy that the ultrametric $u_X(x, x')$ between any two different points $x$ and $x'$ cannot be smaller than the $\alpha$-separation $\sep^\alpha(X, \dl_X, \du_X)$, i.e.
\begin{align}\label{eqn_property_minimum_separation}
    u_X(x, x') \geq \sep^\alpha(X, \dl_X, \du_X) \qquad \forall x \neq x'.
\end{align}\end{mylist}

Equivalently, the output dendrogram is such that for resolutions $\delta < \sep^\alpha(X, \dl_X, \du_X)$, each node is in its own block. We note that (P1) does not requires that a cluster with more than one node is formed at resolution $\sep^\alpha(X, \dl_X, \du_X)$ but states that achieving this minimum resolution is a prerequisite condition for the emergence of clusters. Property of Minimum Separation does not only provide intuition in more complicated scenarios than two-node spaces, but is also substantial for later developments in the paper; see, e.g. the proof of Theorem \ref{thm_bounds}.

Notice that if we apply the Property of Minimum Separation (P1) onto the two-node space $\Delta_2(\dl, \du)$, we must have $u_{p,q}(p,q) \geq \sep_\alpha(\{p, q\}, \dl, \du) = \alpha \du + (1 - \alpha) \dl$, which means that (P1) and the Axiom of Value (A1) are compatible requirements. We can therefore construct two alternative axiomatic formulations where admissible methods are required to satisfy the Axiom of Transformation (A2) as well as (P1), or (A2) as well as (A1). As we demonstrate in the following theorem that (P1) is implied by (A2) and (A1). Therefore, both two formulations are equivalent to requiring the fulfillment of axioms (A1) and (A2).

%
\begin{theorem}\label{thm_value_implies_minimum_separation}
If a hierarchical clustering method satisfies the Stronger Axiom of Value (A1') and Axiom of Transformation (A2), it satisfies the Property of Minimum Separation (P1).
\end{theorem}

%
\begin{myproof}
See Appendix \ref{apx_proof_thm_2}.
\end{myproof}

%
\begin{figure*}[t]
\centerline{

\def \thisplotscale {0.9}

\def \unit {\thisplotscale cm}

\tikzstyle {vertex} = [circle, 
                       draw,
                       minimum width = 0.8*\unit,
                       minimum height = 0.8*\unit,
                       anchor=center,
                       font=\scriptsize]
\tikzstyle {light}  = [opacity = 0.1]
\tikzstyle{bigvertex} = [vertex, 
                        minimum width  = 0.7*\unit,
                        minimum height = 0.7*\unit]

\def \dis {3.2}
\def \disy {0.6}
{\footnotesize \begin{tikzpicture}[x = 1.2*\unit, y = 1.1*\unit]

	\fill [blue!7] (0.0 -2,-\disy - 0.6) rectangle (4 * \dis + 0.6, \disy + 0.6);
	\fill [red!15] (0.0 -0.4,0.0) rectangle (4 * \dis + 0.4, \disy + 0.5);
	\fill [green!9] (0.0 -0.4,0.0) rectangle (4 * \dis + 0.4, -\disy - 0.5);
	
	\path (center1) ++ (0+0.2, \disy + 0.3) node (uu) {\red{$\cu_X(x, x')$}};
	\path (center1) ++ (0+0.2, -\disy - 0.3) node (ul) {\green{$\cl_X(x, x')$}};
	
	\path (center1) ++ (0-1, \disy) node (au) {$\alpha \ \times $};
	\path (center1) ++ (0-1, -\disy) node (al) {$(1 - \alpha) \ \times$};
	\path (center1) ++ (0-1, 0) node (aplus) {$+$};
	\path (center1) ++ (0-1.6, \disy + 0.4) node {\blue{$\ccalH^\SL$}};
        
	\node at (0, 0) (center1) {};
	\path (center1) node (1) [fill = blue!20, vertex] {\small $x$};
	\path (1) ++ ( \dis, \disy) node (2) [fill = blue!20, vertex] {\small $x_1$};
	\path (2) ++ ( \dis, 0) node (3) {\small $\cdot \cdot \cdot \cdot \cdot$};
	\path (3) ++ ( \dis, 0) node (4) [fill = blue!20, vertex] {\small $x_{l-1}$};
	
	\path (1) ++ ( \dis, -\disy) node (2l) [fill = blue!20, vertex] {\small $x_1'$};
	\path (2l) ++ ( \dis, 0) node (3l) {\small $\cdot \cdot \cdot \cdot \cdot$};
	\path (3l) ++ ( \dis, 0) node (4l) [fill = blue!20, vertex] {\small $x_{l-1}'$};
	
	\path (4) ++ ( \dis, -\disy) node (5) [fill = blue!20, vertex] {\small $x'$};
	
	\path[-stealth, -] (1) edge [above] node {$\du_X(x, x_1)$} (2);
	\path[-stealth, -] (2) edge [above] node {$\du_X(x_1, x_2)$} (3);
	\path[-stealth, -] (3) edge [above] node {$\du_X(x_{l-2}, x_{l-1})$} (4);
	\path[-stealth, -] (4) edge [above] node {$\du_X(x_{l-1}, x')$} (5);
	
	\path[-stealth, -] (1) edge [below] node {$\dl_X(x, x_1')$} (2l);
	\path[-stealth, -] (2l) edge [below] node {$\dl_X(x_1', x_2')$} (3l);
	\path[-stealth, -] (3l) edge [below] node {$\dl_X(x_{l-2}', x_{l-1}')$} (4l);
	\path[-stealth, -] (4l) edge [below] node {$\dl_X(x_{l-1}', x')$} (5);
	
\end{tikzpicture}}  }
\caption{Cluster-and-combine clustering. Nodes $x$ and $x'$ are clustered together at resolution $\delta$ if there exists a chain such that the maximum convex combination $\alpha \cu_X(x_i, x_{i+1}) + (1 - \alpha) \cl_X(x_i, x_{i+1})$ of minimum upper and lower chain costs is no greater than $\delta$ [cf. \eqref{eqn_cluster_and_combine}]. Of all methods that satisfy the Axioms of Value and Transformation, cluster-and-combine clustering yields the smallest ultrametric between any pair of nodes.}
\label{fig_cluster_and_combine}
\end{figure*}
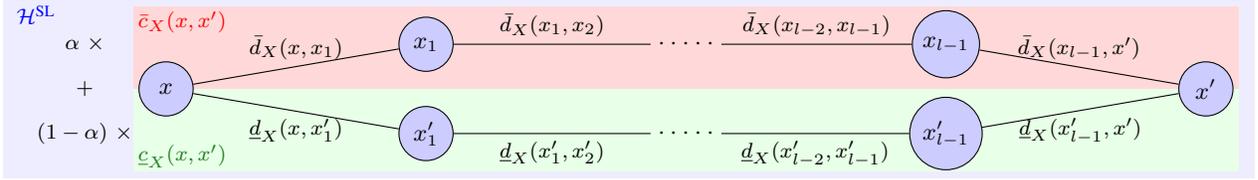

%
\section{Admissible Ultrametrics}\label{sec_admissible_ultrametrics}

Consider a specific metric space with distances given by intervals $I_X = (X, \dl_X, \du_X) \in \ccalI$. Given a value $0\leq\alpha\leq 1$, one particular clustering method satisfying axioms (A1) and (A2) can be established by examining the {\it $\alpha$-combined} dissimilarity
\begin{align}\label{eqn_combine_dissimilarity}
    \hhatd_X(x, x') := \alpha \du_X(x, x') + (1 - \alpha) \dl_X(x, x'),
\end{align}
for any pair of nodes $x, x' \in X$. Though $\hhatd_X$ does not necessarily satisfy the triangle inequality as the original metric distance $d_X$, it is symmetric; therefore the $\alpha$-combined dissimilarity effectively reduces the problem to clustering of symmetric data, a case where the single linkage method defined in \eqref{eqn_single_linkage_ultrametric} is shown to abide to axioms analogous to (A1) and (A2) \cite{carlsson2010}. Based on this observation, we define the {\it combine-and-cluster} method $\ccalH^\CO$ with output $(X, u_X^\CO) = \ccalH^\CO(X, A_X)$ between a pair $x$ and $x'$ as
\begin{align}\label{eqn_combine_and_cluster}
    u_X^\CO(x, x') := \min_{C(x, x')} \max_{i \mid x_i \in C(x, x')} \hhatd_X(x_i, x_{i+1}).
\end{align}
An illustration of the combine-and-cluster clustering method is shown in Fig.~\ref{fig_combine_and_cluster}. For a given pair of nodes $x$ and $x'$, we look for chains $C(x, x')$ connecting them. For a considered chain we examine each of its link, connecting say $x_i$ with $x_{i+1}$, and investigate the convex combination of the distance bounds, i.e. the value of $ \hhatd_X(x_i, x_{i+1}) = \alpha \du_X(x_i, x_{i+1}) + (1 - \alpha) \dl_X(x_i, x_{i+1})$. The maximum value across all links in this chain is then recorded. The combine-and-cluster ultrametric $u_X^\CO(x, x')$ between points $x$ and $x'$ is the minimum of this value across all possible chains connecting $x$ and $x'$. Following the equivalence between dendrograms and ultrametrics as in Theorem \ref{thm_deodemgram_ultrametric}, we can write the combine-and-cluster equivalence classes at resolution $\delta$ as
\begin{align}\label{eqn_combine_and_cluster_dendrogram}
     x \sim_{\CO_X(\delta)} x' \Leftrightarrow
        \min_{C(x, x')} \max_{i \mid x_i \in C(x, x')} \hhatd_X(x_i, x_{i+1}) \leq \delta.
\end{align}
We prove that the output $u_X^\CO$ is a valid ultrametric and the method $\ccalH^\CO$ satisfies axioms (A1) and (A2) as next.

%
\begin{proposition}\label{prop_combine_and_cluster}
Given any value of $0\leq \alpha \leq 1$, the combine-and-cluster method $\ccalH^\CO$ is valid and admissible. I.e., $u_X^\CO$ defined by \eqref{eqn_combine_and_cluster} is an ultrametric for all $I_X = (X, \dl_X, \du_X)$ and $\ccalH^\CO$ satisfies axioms (A1) and (A2).
\end{proposition}

%
\begin{myproof}
One way to see the validity of the ultrametric $u_X^\CO(x, x')$ is because that it is the result of applying single linkage hierarchical clustering on the symmetric dissimilarity $\hhatd_X(x, x')$. Nonetheless, here we give a direct verification. The fact that $u_X^\CO(x, x') = 0$ if and only if $x = x'$ follows directly from that $\hhatd_X(\tdx, \tdx') > 0$ for any distinct nodes which results from the requirements on the bounds $0 < \dl _X(\tdx, \tdx')\leq \du _X(\tdx, \tdx')$. The symmetry property $u_X^\CO(x, x') = u_X^\CO(x', x)$ is because the definition only depends on $\hhatd_X(\tdx, \tdx')$ which is symmetric on $\tdx$ and $\tdx'$. To verify the strong triangle inequality in \eqref{eqn_ultrametrics}, let $C'(x, x')$ and $C''(x', x'')$ be the chains that achieve the minimum in \eqref{eqn_combine_and_cluster} for $u_X^\CO(x, x')$ and $u_X^\CO(x', x'')$, respectively. The maximum convex combination in the concatenated chain $C(x, x'') = C'(x, x') \uplus C''(x', x'')$ does not exceed the maximum cost in each of the individual chains. Therefore, despite that $C(x, x'')$ may not be the optimal chain in \eqref{eqn_combine_and_cluster}, it suffices to bound $u_X^\CO(x, x'') \leq \max \big( u_X^\CO(x, x'), u_X^\CO(x', x'') \big)$ as the strong triangle inequality \eqref{eqn_ultrametrics}.

To see that the Axiom of Value (A1) is satisfied for any considered value $0 \leq \alpha \leq1$, pick an arbitrary two-node space $\Delta_2(\dl, \du)$ and denote $(\{p, q\}, u_{p,q}^\CO) = \ccalH^\CO(\Delta_2(\dl, \du))$ as the output of applying cluster-and-combine clustering method to $\Delta_2(\dl, \du)$. Since every possible chain from $p$ to $q$ must include a link from $p$ to $q$, applying the definition in \eqref{eqn_combine_and_cluster} implies
\begin{align}\label{eqn_combine_and_cluster_A1}
    u_{p,q}^\CO(p, q) = \alpha \du + (1 - \alpha) \dl,
\end{align}
from which axiom (A1) is satisfied.

To verify the fulfillment of axiom (A2), consider $(X, \dl_X, \du_X)$, $(Y, \dl_Y, \du_Y)$, a given value $0 \leq \alpha \leq 1$, and a $\alpha$-distance-reducing map $\phi:X \rightarrow Y$. Let $(X, u_X^\CO) = \ccalH^\CO(X, \dl_X, \du_X)$ and $(Y, u_Y^\CO) = \ccalH^\CO(Y, \dl_Y, \du_Y)$ be the outputs of applying the combine-and-cluster clustering methods onto them. For any pair of nodes $x, x' \in X$, denote $C_X^\star(x, x') = [x = x_0, x_1, \dots, x_{l-1}, x_l = x']$ as the optimal chain in \eqref{eqn_combine_and_cluster} and therefore we can write
\begin{align}\label{eqn_combine_and_cluster_A2_begin}
    u_X^\CO(x, x') = \max_{i \mid x_i \in C^\star_X(x, x')} \hhatd_X(x_i, x_{i+1}).
\end{align}
Consider the mapped chain $C_Y(\phi(x), \phi(x')) = [\phi(x) = \phi(x_0), \dots, \phi(x_l) = \phi(x')]$ in the node space $Y$ under the map $\phi$. Since $\phi$ is $\alpha$-distance-reducing, we have 
\begin{align}\label{eqn_combine_and_cluster_distance_reducing}
    \hhatd_Y(\phi(x_i), \phi(x_{i+1})) \leq \hhatd_X(x_i, x_{i+1}), 
\end{align}
for any $x_i \in C^\star(x, x')$. Combining \eqref{eqn_combine_and_cluster_A2_begin} and \eqref{eqn_combine_and_cluster_distance_reducing} yields
\begin{align}\label{eqn_combine_and_cluster_relation}
    \max_{\phi(x_i) \in C_Y(\phi(x), \phi(x'))}\hhatd_Y(\phi(x_i), \phi(x_{i+1})) \leq u_X^\CO(x, x'). 
\end{align}
Since $C_Y(\phi(x), \phi(x'))$ is a particular chain connecting $\phi(x)$ and $\phi(x')$, the optimal chain cost can only be smaller. Hence,
\begin{align}\label{eqn_combine_and_cluster_optiaml_chain}
    u_Y^\CO(\phi(x), \phi(x')) \!\leq\!\! \max_{\phi(x_i) \in C_Y(\phi(x), \phi(x'))}\! \hhatd_Y(\phi(x_i), \phi(x_{i+1})). 
\end{align}
Finally, substituting \eqref{eqn_combine_and_cluster_optiaml_chain} into \eqref{eqn_combine_and_cluster_relation} demonstrates that $u_Y^\CO(\phi(x), \phi(x')) \leq u_X^\CO(x, x')$, which is the requirement \eqref{eqn_axiom_of_transformation_ultrametric} in the statement of Axiom of Transformation (A2).
\end{myproof}

%
In combine-and-cluster clustering, nodes $x$ and $x'$ belong to the same cluster at resolution $\delta$ whenever we can find a single chain such that the maximum convex combination of distance bounds is no greater than $\delta$. In {\it cluster-and-combine} clustering, we switch the order of operations and investigate chains, potentially different, connecting $x$ and $x'$, with one chain focusing on the distance upper bounds and the other chain examining the distance lower bounds, before combining the upper and lower estimations. To state this definition regarding ultrametrics, consider $I_X = (X, \dl_X, \du_X)$ and  $0\leq \alpha \leq 1$. We define the cluster-and-combine method $\ccalH^\CL$ with output $(X, u_X^\CL) = \ccalH^\CL(X, \dl_X, \du_X)$ as
\begin{equation}\begin{aligned}\label{eqn_cluster_and_combine}
    u_X^\CL(x, x') := & \min_{C(x, x')} \max_{i \mid x_i \in C(x, x')} \Big( \alpha \cu_X(x_i, x_{i+1})  \\
         & ~~~~ + (1 - \alpha) \cl_X(x_i, x_{i+1}) \Big),
\end{aligned}\end{equation}
where recall $\cu_X$ and $\cl_X$ is the minimum upper and lower chain costs defined in \eqref{eqn_min_upper_costs} and \eqref{eqn_min_lower_costs}. An illustration of the cluster-and-combine clustering method is described in Fig.~\ref{fig_cluster_and_combine}. For any pair of nodes, we consider the minimum upper chain cost $\cu_X(x, x')$ as the value $\max_{i \mid x_i \in C(x, x')} \du_X(x_i, x_{i+1})$ fulfilled by the chain $C'(x, x')$ and the minimum lower chain cost $\cl_X(x, x')$ achieved using the chain $C''(x, x')$. The convex combination $\alpha \cu_X(x, x') + (1 - \alpha) \cl_X(x, x')$ is then recorded and the output of the cluster-and-combine clustering method is the result by applying single linkage $\ccalH^\SL$ [cf. \eqref{eqn_single_linkage_ultrametric}]. The single linkage is applied towards $\alpha \cu_X(x, x') + (1 - \alpha) \cl_X(x, x')$ because convex combination of ultrametrics is a metric but not necessarily an ultrametric.  Using the shorthand notation $\hhatc_X(x, x') = \alpha \cu_X(x, x') + (1 - \alpha) \cl_X(x, x')$, the output ultrametric of cluster-and-combine clustering is
\begin{equation}\begin{aligned}\label{eqn_cluster_and_combine_hhatc}
    u_X^\CL(x, x') := & \min_{C(x, x')} \max_{i \mid x_i \in C(x, x')} \hhatc_X(x_i, x_{i+1}).
\end{aligned}\end{equation}
As the case for combine-and-cluster clustering methods, we demonstrate that the output $u_X^\CL$ is a valid ultrametric and the method $\ccalH^\CL$ abides to axioms (A1) and (A2) next.

%
\begin{proposition}\label{prop_cluster_and_combine}
The combine-and-cluster method $\ccalH^\CL$ is valid and admissible. I.e., $u_X^\CL$ defined by \eqref{eqn_cluster_and_combine} is an ultrametric for all $I_X = (X, \dl_X, \du_X)$ and $\ccalH^\CL$ satisfies axioms (A1) and (A2).
\end{proposition}

%
\begin{myproof}
See Appendix \ref{apx_proof_prop_2}.
\end{myproof}

%
\section{Extremal Ultrametrics}\label{sec_extremal_ultrametrics}

Given that we have constructed two admissible methods satisfying axioms (A1)-(A2), it is natural to ask whether these two constructions are the only possible ones, and if not, whether they are special with respect to other satisfying methods. We prove in this section the important characterization that any method $\ccalH$ satisfying axioms (A1)-(A2) yields ultrametrics that lie between $u_X^\CL$ and $u_X^\CO$. The characterization can be considered as a generalization of Theorem 18 in \cite{carlsson2010} for metric spaces.

%
\begin{theorem}\label{thm_bounds}
Consider an admissible clustering method $\ccalH$ satisfying axioms (A1)-(A2). For an arbitrary $I_X = (X, \du_X, \dl_X)$ and $0\leq \alpha \leq 1$, denote $(X, u_X) = \ccalH(I_X)$ the output of applying $\ccalH$ onto $I_X$. Then for any pair of nodes $x, x' \in X$, 
\begin{align}\label{eqn_thm_bounds}
    u_X^\CL(x, x') \leq u_X(x, x') \leq u_X^\CO(x, x'),
\end{align}
where $u_X^\CL(x, x')$ and $u_X^\CO(x, x')$ denote the cluster-and-combine and combine-and-cluster ultrametrics defined in \eqref{eqn_cluster_and_combine} and \eqref{eqn_combine_and_cluster}.
\end{theorem}

%
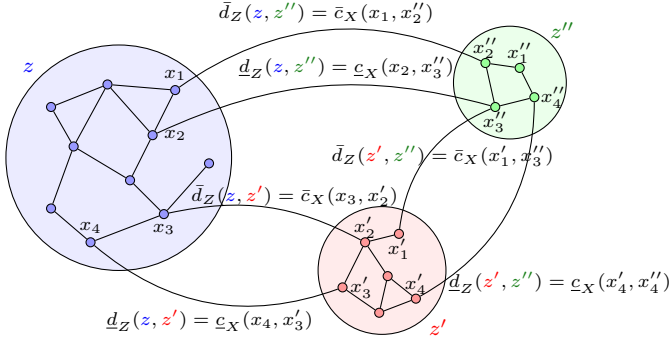
\begin{figure}[t]
\centerline{

\def \thisplotscale {1}

\def \unit {\thisplotscale cm}

\tikzstyle {vertex} = [circle, 
                       draw,
                       minimum width = 0.001*\unit,
                       minimum height = 0.001*\unit,
                       anchor=center]
\tikzstyle {light}  = [opacity = 0.1]
\tikzstyle{bigvertex} = [vertex, 
                        minimum width  = 0.7*\unit,
                        minimum height = 0.7*\unit]

\pgfmathsetmacro{\nodebasesize}{1} 
\pgfmathsetmacro{\nodeinnersep}{0.1}

{\scriptsize \begin{tikzpicture}[x = 1*\unit, y = 1*\unit]

	\node at (0, 0) (center) {};
	\path (center) ++ (0, 0) node (A) [fill = blue!8, vertex, minimum width  = 3*\unit,
                        minimum height = 3*\unit] {};
	\path (center) ++ (3.5, -1.5) node (B) [fill = red!8, vertex, minimum width  = 1.7*\unit,
                        minimum height = 1.7*\unit] {};
	\path (center) ++ (5.2, 1) node (C) [fill = green!8, vertex, minimum width  = 1.5*\unit,
                        minimum height = 1.5*\unit] {};
                        
	\path (A) ++ (-1.2, 1.2) node {\footnotesize \blue{$z$}};   
	
	\path (A) ++ (1.5*0.5, 1.5*0.6) node (a1) [fill = blue!40, point] {};    
	\path (a1) ++ (0, 0.2) node {\scriptsize $x_1$};   
	\path (A) ++ (1.5*0.3, 1.5*0.2) node (a2) [fill = blue!40, point] {};     
	\path (a2) ++ (0.3, 0) node {\scriptsize $x_2$};    
	\path (A) ++ (-1.5*0.1, 1.5*0.65) node (a3) [fill = blue!40, point] {};    
	\path (A) ++ (-1.5*0.4, 1.5*0.1) node (a4) [fill = blue!40, point] {};    
	\path (A) ++ (-1.5*0.6, 1.5*0.45) node (a5) [fill = blue!40, point] {};    
	\path (A) ++ (1.5*0.8, -1.5*0.05) node (a6) [fill = blue!40, point] {};    
	\path (A) ++ (1.5*0.4, -1.5*0.5) node (a7) [fill = blue!40, point] {};    
	\path (a7) ++ (0, -0.2) node {\scriptsize $x_3$};    
	\path (A) ++ (1.5*0.1, -1.5*0.2) node (a8) [fill = blue!40, point] {};    
	\path (A) ++ (-1.5*0.25, -1.5*0.75) node (a9) [fill = blue!40, point] {};    
	\path (a9) ++ (0, 0.2) node {\scriptsize $x_4$};    
	\path (A) ++ (-1.5*0.6, -1.5*0.45) node (a10) [fill = blue!40, point] {};    

	\path (B) ++ (0.75, -0.75) node {\footnotesize \red{$z'$}};   
	
	\path (B) ++ (.75*0.3, .75*0.65) node (b1) [fill = red!40, point] {};     
	\path (b1) ++ (0, -0.2) node {\scriptsize $x_1'$};    
         \path (B) ++ (.75*-0.3, .75*0.5) node (b2) [fill = red!40, point] {};     
	\path (b2) ++ (0, 0.2) node {\scriptsize $x_2'$};    	
	\path (B) ++ (.75*-0.7, .75*-0.3) node (b3) [fill = red!40, point] {};       
	\path (b3) ++ (0.25, 0) node {\scriptsize $x_3'$};    
	\path (B) ++ (.75*0.1, .75*-0.1) node (b4) [fill = red!40, point] {};   
	\path (B) ++ (.75*-0.05, .75*-0.75) node (b5) [fill = red!40, point] {};   
	\path (B) ++ (.75*0.6, .75*-0.5) node (b6) [fill = red!40, point] {};        
	\path (b6) ++ (0, 0.2) node {\scriptsize $x_4'$};   
	
	\path (C) ++ (0.7, 0.7) node {\footnotesize \green{$z''$}};   
	
	\path (C) ++ (.65*0.2, .65*0.3) node (c1) [fill = green!40, point] {}; 
	\path (c1) ++ (0, 0.2) node {\scriptsize $x_1''$};       
	\path (C) ++ (.65*-0.5, .65*0.4) node (c2) [fill = green!40, point] {};   
	\path (c2) ++ (0, 0.2) node {\scriptsize $x_2''$};     
	\path (C) ++ (.65*-0.3, .65*-0.5) node (c3) [fill = green!40, point] {}; 
	\path (c3) ++ (0, -0.2) node {\scriptsize $x_3''$};      
	\path (C) ++ (.65*0.5, .65*-0.3) node (c4) [fill = green!40, point] {};  
	\path (c4) ++ (0.25, 0) node {\scriptsize $x_4''$};    
    
	\path[-stealth, -] (a1) edge [] node {} (a2);
	\path[-stealth, -] (a1) edge [] node {} (a3);
	\path[-stealth, -] (a2) edge [] node {} (a3);
	\path[-stealth, -] (a2) edge [] node {} (a8);
	\path[-stealth, -] (a3) edge [] node {} (a4);
	\path[-stealth, -] (a3) edge [] node {} (a5);
	\path[-stealth, -] (a4) edge [] node {} (a5);
	\path[-stealth, -] (a4) edge [] node {} (a10);
	\path[-stealth, -] (a4) edge [] node {} (a8);
	\path[-stealth, -] (a10) edge [] node {} (a9);
	\path[-stealth, -] (a8) edge [] node {} (a7);
	\path[-stealth, -] (a7) edge [] node {} (a6);
	\path[-stealth, -] (a7) edge [] node {} (a9);

	\path[-stealth, -] (b1) edge [] node {} (b2);
	\path[-stealth, -] (b2) edge [] node {} (b4);
	\path[-stealth, -] (b2) edge [] node {} (b3);
	\path[-stealth, -] (b3) edge [] node {} (b5);
	\path[-stealth, -] (b4) edge [] node {} (b5);
	\path[-stealth, -] (b4) edge [] node {} (b6);
	\path[-stealth, -] (b5) edge [] node {} (b6);
	
	\path[-stealth, -] (c1) edge [] node {} (c2);
	\path[-stealth, -] (c2) edge [] node {} (c3);
	\path[-stealth, -] (c3) edge [] node {} (c4);
	\path[-stealth, -] (c4) edge [] node {} (c1);
	
	\path[-stealth, -] (a7) edge [bend left = 20, above, pos = 0.65] node {$\du_Z(\blue z, \red{z'}) = \cu_X(x_3, x_2')$} (b2);
	\path[-stealth, -] (a9) edge [bend right, below] node {$\dl_Z(\blue z, \red{z'}) = \cl_X(x_4, x_3')$} (b3);
	
	\path[-stealth, -] (a1) edge [bend left, above] node {$\du_Z(\blue z, \green{z''}) = \cu_X(x_1, x_2'')$} (c2);
	\path[-stealth, -] (a2) edge [bend left = 20, above, pos = 0.58] node {$\dl_Z(\blue z, \green{z''}) = \cl_X(x_2, x_3'')$} (c3);

	\path[-stealth, -] (b1) edge [bend left, below, pos = 0.65] node {$\du_Z(\red{z'}, \green{z''}) = \cu_X(x_1', x_3'')$} (c3);
	\path[-stealth, -] (b6) edge [bend right, right, pos = 0.1] node {$\dl_Z(\red{z'}, \green{z''}) = \cl_X(x_4', x_4'')$} (c4);
\end{tikzpicture}} }
\caption{Network of equivalence classes at a given resolution. Each shaded subset of nodes represent an equivalence class. The Axiom of Transformation establishes the relationship between the clustering of nodes in the original network and the clustering of nodes in the network of equivalence classes.}
\label{fig_equivalence_class}
\end{figure}

%
\begin{myproof}[of $\mathbf{u_X^\CL(x, x') \leq u_X(x, x')}$]
By Theorem \ref{thm_value_implies_minimum_separation}, $\ccalH$ satisfying (A1)-(A2) implies that it also satisfies (P1). To show the first inequality in \eqref{eqn_thm_bounds}, consider the cluster-and-combine clustering equivalence relation $\sim_{\CL_X(\delta)}$ at resolution $\delta$ using $x \sim_{\CL_X(\delta)} x'$  if and only if $u_X^\CL(x, x') \leq \delta$. Define the space $Z := X \mod \sim_{\CL_X(\delta)}$ where points in $X$ belonging to the same equivalence class are represented by a single node in $Z$ and the map $\phi_\delta: X \rightarrow Z$ that maps each point of $X$ to its equivalence class. Points $x$ and $x'$ are mapped to the same point under $\phi_\delta$ if and only if they belong to the same equivalence class at $\delta$, i.e.
\begin{align}\label{eqn_proof_thm_bounds_1_Z}
    \phi_\delta(x) = \phi_\delta(x') ~ \Longleftrightarrow ~ u_X^\CL(x, x') \leq \delta.
\end{align}
We define the metric space with distances given by intervals $I_Z := (Z, \du_Z, \dl_Z)$ by equipping $Z$ with distance bounds as
\begin{equation}\begin{aligned}\label{eqn_proof_thm_bounds_1_dZ}
    \dl_Z(z, z') &:= \min_{x \in \phi_\delta^{-1}(z), x' \in \phi_\delta^{-1}(z')} \cl_X(x, x'), \\
\end{aligned}\end{equation}
and similarly for $\du_Z(z, z')$. The distance lower bounds $\dl_Z(z, z')$ compares all the minimum lower chain costs $\cl_X(x, x')$ between a member of the equivalence class $z$ and a member of the equivalence class $z'$ and sets $\dl_Z(z, z')$ to the value corresponding to the pair yielding the lowest minimum lower chain cost. The distance upper bounds $\du_Z(z, z')$ are constructed similarly; see Fig.~\ref{fig_equivalence_class}. Observe that follows from the construction, the map $\phi_\delta$ is $\alpha$-distance-reducing such that for any $x, x' \in X$
\begin{align}\label{eqn_proof_thm_bounds_1_distance_reducing_d}
    \hhatd_X(x, x') &\geq \hhatd_Z(\phi_\delta(x), \phi_\delta(x')), \\
        \label{eqn_proof_thm_bounds_1_distance_reducing_c}
            \hhatc_X(x, x') &\geq \hhatc_Z(\phi_\delta(x), \phi_\delta(x')).
\end{align}
To see this, when $x$ and $x'$ are co-clustered at resolution $\delta$, $\hhatd_Z(\phi_\delta(x), \phi_\delta(x')) = \hhatc_Z(\phi_\delta(x), \phi_\delta(x')) = 0$. Otherwise, if they are mapped to different equivalent classes, we can write
\begin{equation}\begin{aligned}\label{eqn_proof_thm_bounds_1_distance_reducing_d_1}
    \dl_X(x, x') \geq \cl_X(x, x') &\geq \min_{x \in \phi_\delta^{-1}(z), x' \in \phi_\delta^{-1}(z')} \cl_X(x, x') \\
        & = \dl_Z(\phi_\delta(x), \phi_\delta(x')),
\end{aligned}\end{equation}
and similarly $\du_X(x, x') \geq \du_Z(\phi_\delta(x), \phi_\delta(x'))$. Eq. \eqref{eqn_proof_thm_bounds_1_distance_reducing_d} then follows from these two inequalities. Besides, we can also write
\begin{equation}\begin{aligned}\label{eqn_proof_thm_bounds_1_distance_reducing_c_1}
    \cl_X(x, x') &\geq \min_{x \in \phi_\delta^{-1}(z), x' \in \phi_\delta^{-1}(z')} \cl_X(x, x') \\
        & = \dl_Z(\phi_\delta(x), \phi_\delta(x')) \geq \cl_Z(\phi_\delta(x), \phi_\delta(x')),
\end{aligned}\end{equation}
and similarly $\cu_X(x, x') \geq \cu_Z(\phi_\delta(x), \phi_\delta(x'))$. The convex combination of these two inequalities is identical to Eq. \eqref{eqn_proof_thm_bounds_1_distance_reducing_c}. This completes the proof that $\phi_\delta$ is $\alpha$-distance-reducing.

Consider a clustering method $\ccalH$ satisfying axioms (A1)-(A2) and write $(Z, u_Z) = \ccalH(I_Z)$ as the output of applying $\ccalH$ upon $I_Z$. To apply (P1) we investigate the $\alpha$-separation of $I_Z$ as next.

%
\begin{fact}\label{fact_alpha_separation}
The $\alpha$-separation of $I_Z$ is
\begin{align}\label{eqn_fact_alpha_separation}
    \textmd{\sep}^\alpha(I_Z) > \delta.
\end{align}\end{fact}

%
\begin{myproof}
Suppose the contrary is true, i.e. $\sep^\alpha(I_Z) \leq \delta$, then there exists a pair of distinct nodes $z \neq z' \in Z$ such that the convex combination of their distance bounds satisfies
\begin{equation}\begin{aligned}\label{eqn_fact_alpha_separation_proof_1}
    &\alpha \min_{C(z, z')} \max_{i \mid z_i \in C(z, z')} \du_Z(z_i, z_{i+1}) \\
        &\quad+(1 - \alpha) \min_{\tdC(z, z')} \max_{i \mid z_i \in \tdC(z, z')} \dl_Z(z_i, z_{i+1}) \leq \delta.
\end{aligned}\end{equation}
Denote $C^\star$ as the optimal chain in minimizing $\min_{C(z, z')}  \max_{i \mid z_i \in C(z, z')} \du_Z(z_i, z_{i+1})$ and $\tdC^\star$ as the chain in $\min_{\tdC(z, z')} \max_{i \mid z_i \in \tdC(z, z')} \dl_Z(z_i, z_{i+1})$, \eqref{eqn_fact_alpha_separation_proof_1} then becomes
\begin{equation}\begin{aligned}\label{eqn_fact_alpha_separation_proof_2}
    \alpha \!\!\!\! \max_{i \mid z_i \in C^\star(z, z')}\!\!\! \du_Z(z_i, \!z_{i+1}) \!+\! (1 \!-\! \alpha) \!\!\!\! \max_{i \mid z_i \in \tdC^\star(z, z')}\!\!\! \dl_Z(z_i, \!z_{i+1}) \!\leq\! \delta.
\end{aligned}\end{equation}

From the definitions of $\dl_Z$ given by \eqref{eqn_proof_thm_bounds_1_dZ} and $\du_Z$, we can find four nodes $x, \tdx, x', \tdx'$ with $\phi_\delta(x) = \phi_\delta(\tdx) = z$, $\phi_\delta(x') = \phi_\delta(\tdx') = z'$, and two chains $C^\star(x, x')$ and $\tdC^\star(\tdx, \tdx')$ which are mapped to $C^\star(z, z')$ and $\tdC^\star(z, z')$ under $\phi_\delta$ such that
\begin{equation}\begin{aligned}\label{eqn_fact_alpha_separation_proof_3}
    \alpha \!\! \max_{i \mid x_i \in C^\star(x, x')}\!  \cu_X(x_i, \!x_{i+1}) \!+\!(1 \!-\! \alpha) \!\! \max_{i \mid x_i \in \tdC^\star(\tdx, \tdx')} \! \cl_X(x_i, x_{i+1}) \!\leq\! \delta.
\end{aligned}\end{equation}
Because $\cu_X$ is a valid ultrametric, we can write
\begin{equation}\begin{aligned}\label{eqn_fact_alpha_separation_proof_ultrametric}
    &\cu_X(x, x') \leq \max\left\{ \cu_X(x, x_1), \cu_X(x_1, x') \right\} \\
        &\quad \leq  \max\left\{ \cu_X(x, x_1), \cu_X(x_1, x_2), \cu_X(x_2, x') \right\} \\
        &\quad \leq \dots \leq \max_{i \mid x_i \in C^\star(x, x')} \cu_X(x_i, x_{i+1}).
\end{aligned}\end{equation}
Similarly $\cl_X(\tdx, \tdx') \leq \max_{i \mid x_i \in \tdC^\star(\tdx, \tdx')} \cl_X(x_i, x_{i+1})$. Substituting these two bounds into \eqref{eqn_fact_alpha_separation_proof_3} implies
\begin{equation}\begin{aligned}\label{eqn_fact_alpha_separation_proof_4}
    \alpha \cu_X(x, x') + (1 - \alpha) \cl_X(\tdx, \tdx') \leq \delta.
\end{aligned}\end{equation}
Further observe that because $x$ and $\tdx$ belong to the same cluster ($z$) as well as $x'$ and $\tdx'$ belong to the same cluster ($z'$) at resolution $\delta$, we know that $\alpha \cu_X(x, x') + (1 - \alpha) \cl_X(x, x') \leq \delta$ and $\alpha \cu_X(\tdx, \tdx') + (1 - \alpha) \cl_X(\tdx, \tdx') \leq \delta$. To reach a contradiction we use the following fact.

%
\begin{fact}\label{fact_at_least_one}
There exists a pair $\dotx \in \{x, \tdx\}$ and $\dotx' \in \{x', \tdx'\}$ such that
\begin{align}\label{eqn_fact_at_least_one}
    \alpha \cu_X(\dotx, \dotx') + (1 - \alpha) \cl_X(\dotx, \dotx') \leq \delta.
\end{align}\end{fact}

%
\begin{myproof}
Define the following shorthand notations, $\au := \cu_X(x, \tdx)$, $\bu := \cu_X(x', \tdx')$, $\eu := \cu_X(x, x')$, $\fu := \cu_X(\tdx, \tdx')$, $\gu := \cu_X(x, \tdx')$, $\hu := \cu_X(x', \tdx)$. Similarly define $\al, \bl, \el, \fl, \gl, \hl$; see Fig.~\ref{fig_proof_fact_at_least_one}. The problem then becomes: given $\alpha \au + (1 - \alpha) \al \leq \delta$, $\alpha \bu + (1 - \alpha) \bl \leq \delta$, and $\alpha \eu + (1 - \alpha) \fl \leq \delta$, we would like to prove that at least one of the following holds: $\alpha \eu + (1 - \alpha) \el \leq \delta$, $\alpha \fu + (1 - \alpha) \fl \leq \delta$, $\alpha \gu + (1 - \alpha) \gl \leq \delta$, or $\alpha \hu + (1 - \alpha) \hl \leq \delta$. We show the fact by examining which is the maximum one out of $\au, \bu, \eu$ and which is the maximum one out of $\al, \bl, \fl$. 

Firstly, in scenarios where $\eu = \max\{\au, \bu, \eu\}$, it follows from the strong triangle inequality of $\cu_X$ that $\fu \leq \eu$. Therefore,
\begin{align}\label{eqn_proof_fact_at_least_one_firstly}
    \alpha \fu + (1 - \alpha) \fl  \leq \alpha \eu + (1 - \alpha) \fl \leq \delta,
\end{align}
which shows the desired result. The proof for cases with $\fl = \max\{\al, \bl, \fl\}$ follows by symmetry. Therefore, what remain are scenarios where neither $\eu = \max\{\au, \bu, \eu\}$ nor $\fl = \max\{\al, \bl, \fl\}$.

Secondly, in scenarios where $\au = \max\{\au, \bu, \eu\}$ and $\al = \max\{\al, \bl, \fl\}$, we have $\eu \leq \au$ and $\el \leq \al$ where the latter follows from the strong triangle inequality of $\cu_X$. Consequently
\begin{align}\label{eqn_proof_fact_at_least_one_secondly}
    \alpha \eu + (1 - \alpha) \el  \leq \alpha \au + (1 - \alpha) \al \leq \delta,
\end{align}
which shows the desired result. The proof for cases with $\bu = \max\{\au, \bu, \eu\}$ and $\bl = \max\{\al, \bl, \fl\}$ follows by symmetry.

Thirdly, consider $\au = \max\{\au, \bu, \eu\}$ and $\bl = \max\{\al, \bl, \fl\}$. If $\eu \leq \bu$, because $\el \leq \bl$ by the strong inequality of $\cl_X$, we have
\begin{align}\label{eqn_proof_fact_at_least_one_thirdly_1}
    \alpha \eu + (1 - \alpha) \el  \leq \alpha \bu + (1 - \alpha) \bl \leq \delta,
\end{align}
which is the desired result. Otherwise, if $\bu \leq \eu$, we can write $\gu \leq \max\{\bu, \eu\} = \eu$ and $\gl \leq \max\{\al, \fl\}$. Therefore,
\begin{equation}\begin{aligned}\label{eqn_proof_fact_at_least_one_thirdly_2}
    \alpha \gu + (1 - \alpha) \gl \leq \alpha \eu + (1 - \alpha) \max\{\al, \fl \}.
\end{aligned}\end{equation}
Utilizing the fact $\eu \leq \au$ in \eqref{eqn_proof_fact_at_least_one_thirdly_2} yields
\begin{equation}\begin{aligned}\label{eqn_proof_fact_at_least_one_thirdly_2_final}
    \alpha \gu \!+\! (1 \!-\! \alpha) \gl \!\leq \!
        \max\left\{ \alpha \au \!+\! (1 \!-\! \alpha) \al, \alpha \eu \!+\! (1 \!-\! \alpha) \fl \right\} \!\leq\! \delta,
\end{aligned}\end{equation}
which shows the desired result. The proof for $\bu = \max\{\au, \bu, \eu\}$ and $\al = \max\{\al, \bl, \fl\}$ follows by symmetry. We have proven the statement under all cases, and the proof of Fact \ref{fact_at_least_one} is complete.
\end{myproof}

%
Continuing with the proof of Fact \ref{fact_alpha_separation}, since there exists a pair of nodes $\dotx \in \{x, \tdx\}$ and $\dotx' \in \{x', \tdx'\}$ with $\alpha \cu_X(\dotx, \dotx') + (1 - \alpha) \cl_X(\dotx, \dotx') \leq \delta$, the fact $u_X^\CL(x, x') \leq \delta$ contradicts the assumption $\phi(\dotx) = z \neq z' = \phi(\dotx')$. Therefore, the assumption that \eqref{eqn_fact_alpha_separation} is false cannot hold. The opposite must be true.
\end{myproof}

%
Finally, back to the main proof of $u_X^\CL(x, x') \leq u_X(x, x')$, recall that $(Z, u_Z) = \ccalH(Z, \dl_Z, \du_Z)$. Since the $\alpha$-separation of $Z$ satisfies \eqref{eqn_fact_alpha_separation}, (P1) implies for any pair of nodes $z \neq z'$, $u_Z(z, z') > \delta$. Also observe that because $\phi$ is $\alpha$-distance-reducing and $\ccalH$ satisfies (A2), we must have $u_X(x, x') \geq u_Z(z, z')$. This inequality, combined with $u_Z(z, z') > \delta$ enables us to conclude that when $x$ and $x'$ are mapped to different equivalence classes, 
\begin{equation}\begin{aligned}\label{eqn_proof_thm_bounds_1_final_2}
   u_X(x, x') \geq u_Z(z, z') > \delta.
\end{aligned}\end{equation}
Notice that from \eqref{eqn_proof_thm_bounds_1_Z}, $x$ and $x'$ are mapped to different equivalence classes if and only if $u_X^\CL(x, x') > \delta$. Therefore, we can claim that $u_X^\CL(x, x') > \delta$ implies $u_X(x, x') > \delta$. Because this statement is true for any $\delta > 0$, it induces that $u_X^\CL(x, x') \leq u_X(x, x')$ for any $x \neq x' \in X$ as the first inequality in \eqref{eqn_thm_bounds}.
\end{myproof}

%
\begin{figure}[t]
{\center

\def \thisplotscale {0.8}

\def \unit {\thisplotscale cm}

\tikzstyle {vertex} = [circle, 
                       draw,
                       minimum width = 0.05*\unit,
                       minimum height = 0.05*\unit,
                       anchor=center]
\tikzstyle {light}  = [opacity = 0.1]
\tikzstyle{point2} = [point, 
                        minimum width  = 0.5*\unit,
                        minimum height = 0.5*\unit]

\pgfmathsetmacro{\nodebasesize}{1} 
\pgfmathsetmacro{\nodeinnersep}{0.1}

{\footnotesize \begin{tikzpicture}[x = 1.4*\unit, y = 0.8*\unit]

	\node at (0, 0) (center) {};
	\path (center) ++ (0, 0) node (A) [fill = blue!10, vertex, minimum width  = 2.5*\unit,
                        minimum height = 2.5*\unit] {};
	\path (center) ++ (5, 0) node (B) [fill = red!10, vertex, minimum width  = 2.5*\unit,
                        minimum height = 2.5*\unit] {};
                        
	\path (A) ++ (-1.2, 1.2) node {$\blue z$};
	\path (B) ++ (1.22, 1.2) node {$\red{z'}$};
                        
	\path (A) ++ (0, 0.9) node (a1) [fill = white, fill = blue!35, point2] {$ x$};   
	\path (A) ++ (0, -0.9) node (a2) [fill = white, fill = blue!35, point2] {$ \tdx$};  
	
	\path (B) ++ (0, 0.9) node (b1) [fill = white, fill = red!35, point2] {$ x'$};   
	\path (B) ++ (0, -0.9) node (b2) [fill = white, fill = red!35, point2] {$ \tdx'$};  
	
	\path[-stealth, -, left] (a1) edge [] node {$\au$} (a2);
	\path[-stealth, -, line width = 0pt, right] (a1) edge [] node {$\al$} (a2);
	\path[-stealth, bend left = 14, -, above] (a1) edge [] node {$\eu$} (b1);
	\path[-stealth, -, bend left = 14, line width = 0pt, below] (a1) edge [] node {$\el$} (b1);
	\path[-stealth, -, right] (b1) edge [] node {$\bu$} (b2);
	\path[-stealth, -, line width = 0pt, left] (b1) edge [] node {$\bl$} (b2);
	\path[-stealth, bend left = 14, below, -] (b2) edge [] node {$\fu$} (a2);
	\path[-stealth, bend left = 14, above, line width = 0pt, -] (b2) edge [] node {$\fl$} (a2);
	
	\path[-stealth, -, above, pos = 0.2] (a1) edge [] node {$\gu$} (b2);
	\path[-stealth, -, line width = 0pt, below, pos = 0.2] (a1) edge [] node {$\gl$} (b2);
	\path[-stealth, -, above, pos = 0.81] (a2) edge [] node {$\hu$} (b1);
	\path[-stealth, -, line width = 0pt, below, pos = 0.81] (a2) edge [] node {$\hl$} (b1);

\end{tikzpicture}}}
\caption{The illustration in the proof of Fact \ref{fact_at_least_one}. The alphabets adjacent to edges denote the corresponding minimum upper and lower chain costs. Given that $\alpha\au + (1 - \alpha)\al \delta$, $\alpha \bu + (1 - \alpha)\bl \leq \delta$, and $\alpha\eu + (1 - \alpha)\fl \leq \delta$, there exists a pair of nodes $\dotx \in \{x, \tdx\}$ and $\dotx' \in \{x', \tdx'\}$ such that $\alpha \cu_X(\dotx, \dotx') + (1 - \alpha) \cl_X(\dotx, \dotx') \leq \delta.$} 
\label{fig_proof_fact_at_least_one}
\end{figure}
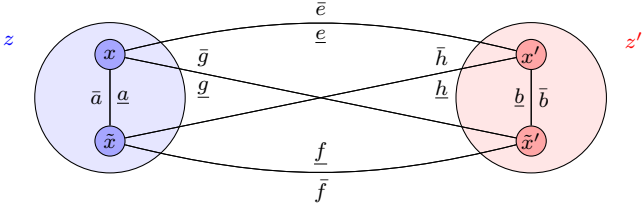

%
\begin{myproof}[of $\mathbf{u_X(x, x') \leq u_X^\CO(x, x')}$]
To show the second inequality in \eqref{eqn_thm_bounds}, first notice that for any distinct nodes $x_i \neq x_j \in X$, we can construct a two-node space $\Delta_{i,j} = \big(\{p, q\}, \dl_X(x_i, x_j), \du_X(x_i, x_j)\big)$ and a mapping $\phi_{i,j} : \{p, q\} \rightarrow X$ with $\phi_{i,j}(p) = x_i$ and $\phi_{i,j}(q) = x_j$ such that $\phi_{i,j}$ is $\alpha$-distance-reducing. To demonstrate this, we need to verify conditions \eqref{eqn_distance_reducing_d} and $\eqref{eqn_distance_reducing_c}$. Eq. \eqref{eqn_distance_reducing_d} follows because $\du_X(\phi_{i,j}(p), \phi_{i,j}(q)) = \du_{p,q}(p, q),\dl_X(\phi_{i,j}(p), \phi_{i,j}(q)) = \dl_{p,q}(p, q)$ and therefore the convex combination of the distance bounds also coincide. To see \eqref{eqn_distance_reducing_c}, using the relationships between distance bounds and minimum chain costs, we can write
\begin{equation}\begin{aligned}\label{eqn_proof_thm_bounds_2_phi_c}
    \cu_X(\phi_{i,j}(p), \phi_{i,j}(q)) &\leq \du_X(\phi_{i,j}(p), \phi_{i,j}(q)), \\
    \cl_X(\phi_{i,j}(p), \phi_{i,j}(q)) &\leq \dl_X(\phi_{i,j}(p), \phi_{i,j}(q)), \\
    \dl_{p,q}(p, q) = \cl_{p,q}(p, q), & \quad \du_{p,q}(p, q) = \cu_{p,q}(p, q).
\end{aligned}\end{equation}
Therefore, have
\begin{equation}\begin{aligned}\label{eqn_proof_thm_bounds_2_phi_c_final}
    \cu_X(\phi_{i,j}(p), \phi_{i,j}(q))\!\leq\! \cu_{p,q}(p, q), ~\cl_X(\phi_{i,j}(p), \phi_{i,j}(q)) \!\leq\! \cl_{p,q}(p, q),
\end{aligned}\end{equation}
from which the requirement of convex combination in \eqref{eqn_distance_reducing_c} follows directly. Because $\ccalH$ satisfies (A1), the output ultrametric $(\{p, q\}, u_{p,q})$ of applying $\ccalH$ onto $\Delta_{i,j}$ implies 
\begin{equation}\begin{aligned}\label{eqn_proof_thm_bounds_2_A1}
    u_{p,q} (p,q) \!=\! \alpha  \du_{p,q}(p, q) + (1 - \alpha) \dl_{p,q}(p,q) \!=\! \hhatd_X(x_i, x_j),
\end{aligned}\end{equation}
Moreover, $\ccalH$ satisfies (A2), and therefore
\begin{equation}\begin{aligned}\label{eqn_proof_thm_bounds_2_A2}
    u_X(x_i, x_j) \leq u_{p,q} (p,q) = \hhatd_X(x_i, x_j).
\end{aligned}\end{equation}
Observe that when $x_i = x_j$, \eqref{eqn_proof_thm_bounds_2_A2} also holds because both sides on the inequality is zero. Consequently, \eqref{eqn_proof_thm_bounds_2_A2} holds true for any points $x_i, x_j \in X$. Now, consider the nodes $x$ and $x'$ and denote $C^\star(x, x')$ as the chain yielding the minimum cost in \eqref{eqn_combine_and_cluster}, 
\begin{equation}\begin{aligned}\label{eqn_proof_thm_bounds_2_combine_and_cluster}
    u_X^\CO(x, x') = \max_{i \mid x_i \in C^\star(x, x')} \hhatd_X(x_i, x_{i+1}).
\end{aligned}\end{equation}
Substituting the inequality \eqref{eqn_proof_thm_bounds_2_A2} in \eqref{eqn_proof_thm_bounds_2_combine_and_cluster} yields
\begin{equation}\begin{aligned}\label{eqn_proof_thm_bounds_2_combine_and_cluster_bounds}
    u_X^\CO(x, x') \geq \max_{i \mid x_i \in C^\star(x, x')} u_X(x_i, x_{i+1}).
\end{aligned}\end{equation}
Finally, because $u_X$ is a valid ultrametric, as in \eqref{eqn_fact_alpha_separation_proof_ultrametric}, we can bound $u_X(x, x') \leq \max_{i \mid x_i \in C^\star(x, x')} u_X(x_i, x_{i+1})$. Combining with \eqref{eqn_proof_thm_bounds_2_combine_and_cluster_bounds} yields $u_X^\CO(x, x') \geq u_X(x, x')$ as the second inequality in \eqref{eqn_thm_bounds}.
\end{myproof}

%
From Theorem \ref{thm_bounds}, cluster-and-combine clustering $u_X^\CL$ applied to $I_X = (X, \dl_X, \du_X)$ yields a minimal ultrametric among outputs by all methods satisfying axioms (A1)-(A2). Combine-and-cluster clustering $u_X^\CO$ yields a uniformly maximal ultrametric.

\begin{remark} \normalfont Theorem \ref{thm_bounds} resembles the results obtained for asymmetric clustering in \cite{carlsson2014} in which two methods are obtained and shown to be extremal with respect to similar axioms. The difference is that in metric spaces with distances given by intervals both upper and lower bounds represent the uncertain but symmetric relationship between the pair. In asymmetric networks, all observations are certain but the relationship from node $x$ to $x'$ is asymmetric and may not be the same as the relationship from node $x'$ to $x$. These differences manifest on the selection of a different axiom of value where instead of clustering at the convex combination $\alpha \du + (1 - \alpha) \dl$ [cf. \eqref{eqn_axiom_of_value_ultrametric}] we cluster at the larger of the two dissimilarities. It follows that combine-and-cluster and cluster-and-combine are admissible and extremal with respect to a different set of axioms than the methods in \cite{carlsson2014}.  \end{remark}

%
\subsection{Hierarchical Clustering Given Extremal Confidence Level}\label{sec_alpha_0_1}

In the previous section, we consider admissible clustering methods given an arbitrary value $0 \leq \alpha \leq 1$. In this subsection, we investigate the special cases given $\alpha$ at the extreme points, i.e. $\alpha \in \{0, 1\}$. Starting with $\alpha = 1$, this means we are the most conservative and believe the distance between two points $x$ and $x'$ being their distance upper bound $\du_X(x, x')$. The output of the combine-and-cluster clustering methods can then be written as
\begin{align}\label{eqn_combine_and_cluster_alpha_1}
    u_X^\CO(x, x') = \min_{C(x, x')} \max_{i \mid x_i \in C(x, x')} \du_X(x_i, x_{i+1}),
\end{align}
which is the same as applying single linkage clustering $\ccalH^\SL$ [cf. \eqref{eqn_single_linkage_ultrametric}] onto distance upper bounds $\du_X$. On the other hand, the output of the cluster-and-combine clustering methods is
\begin{align}\label{eqn_cluster_and_combine_alpha_1}
    u_X^\CL(x, x') = \min_{C(x, x')} \max_{i \mid x_i \in C(x, x')} \cu_X(x_i, x_{i+1}),
\end{align}
with $\cu_X$ the minimum upper chain costs defined in \eqref{eqn_min_upper_costs}. Notice that $\cu_X$ is also the output of applying single linkage clustering $\ccalH^\SL$ onto the distance upper bounds. Moreover, because $\cu_X$ is a valid ultrametric, $\min_{C(x, x')} \max_{i \mid x_i \in C(x, x')} \cu_X(x_i, x_{i+1})$ is the same as $\cu_X(x, x')$. Combining these observations, it follows that
\begin{align}\label{eqn_all_equal_alpha_1}
    u_X^\CL(x, x') = u_X^\SL(x, x') = u_X^\CO(x, x').
\end{align}

When $\alpha = 0$, meaning that we are the most liberate and believe the distance between two points $x$ and $x'$ being their distance lower bound $\dl_X(x, x')$, a similar analysis would follow. We can now utilize Theorem \ref{thm_bounds} and \eqref{eqn_all_equal_alpha_1} to prove the uniqueness of admissible hierarchical clustering methods abiding (A1)-(A2), given that the confidence level $\alpha$ is at the extremes, i.e. $\alpha \in \{0, 1\}$.

%
\begin{corollary}\label{coro_alpha_0_1}
Consider a clustering method $\ccalH$ satisfying axioms (A1)-(A2). For arbitrary $I_X = (X, \du_X, \dl_X)$, denote $(X, u_X) = \ccalH(I_X)$ the output of applying $\ccalH$ onto $I_X$. When $\alpha = 1$, $\ccalH \equiv \ccalH^\SL(\du_X)$ is the same as the singe linkage clustering [cf. \eqref{eqn_single_linkage_ultrametric}] onto the distance upper bounds; when $\alpha = 0$, $\ccalH \equiv \ccalH^\SL(\dl_X)$.
\end{corollary}

%
\begin{myproof}
When $\alpha = 1$, because $\ccalH$ satisfies the hypotheses of Theorem \ref{thm_bounds}, \eqref{eqn_thm_bounds} is true for any distinct nodes $x, x' \in X$. But by \eqref{eqn_all_equal_alpha_1}, cluster-and-combine and combine-and-cluster ultrametrics coincide; as a result \eqref{eqn_thm_bounds} can be written as
\begin{align}\label{eqn_proof_coro_alpha_0_1}
    u_X^\SL(x, x') \leq u_X(x, x') \leq u_X^\SL(x, x').
\end{align}
It follows that $u_X(x, x') = u_X^\SL(x, x')$ for any pair of nodes $x, x'$. Therefore $\ccalH \equiv \ccalH^\SL(\du_X)$. Similar derivation holds for $\alpha = 0$.
\end{myproof}

%
Restrict attention to the metric space $\ccalM \subset \ccalI$ of the form $(X, d_X)$. A further application of Corollary \ref{coro_alpha_0_1} implies singe linkage clustering is the unique admissible methods as next.

%
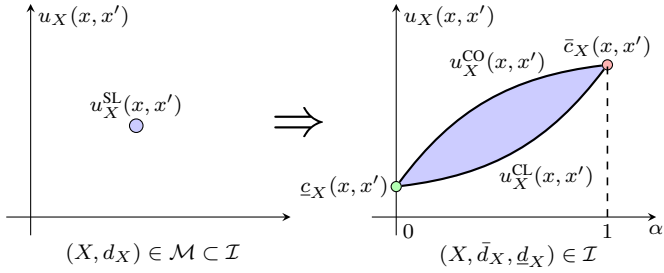
\begin{figure}[t]
\centerline{

\def \thisplotscale {0.9}

\def \unit {\thisplotscale cm}

\tikzstyle {vertex} = [circle, 
                       draw,
                       minimum width = 0.7*\unit,
                       minimum height = 0.7*\unit,
                       anchor=center,
                       font=\small]
\tikzstyle {light}  = [opacity = 0.1]
\tikzstyle{bigvertex} = [vertex, 
                        minimum width  = 0.7*\unit,
                        minimum height = 0.7*\unit]


{\footnotesize \begin{tikzpicture}[x = 1.2*\unit, y = 0.9*\unit]
  
   
    \draw [-stealth] (-0.3,0) -- (3.2,0);
    \draw [-stealth] (0,-0.3) -- (0,3.5);
    \node [right] at (0, 3.3) {$u_X(x, x')$};
    \node [below] at (1.5, -0.3) {$(X, d_X) \in \ccalM \subset \ccalI$};
    
    \node at (1.3, 1.5) [fill = blue!20, point, minimum width  = 0.2*\unit, minimum height = 0.2*\unit] (m1) {};
    \node [above] at (m1) {$u_X^\SL(x, x')$};
    
   
    \draw [-stealth] (4.5-0.3,0) -- (4.5+3.2,0) node [below, at end] {$\alpha$};
    \draw [-stealth] (4.5,-0.3) -- (4.5,3.5);
    \node [right] at (4.5, 3.3) {$u_X(x, x')$};
    \node [below] at (4.5+1.5, -0.3) {$(X, \du_X, \dl_X) \in \ccalI$};
    
    \path [fill = blue!20] (4.5, 0.5) to [bend right = 25] (4.5+2.6, 2.5) to [bend right = 25] (4.5, 0.5);
    \node at (4.5+0, 0.5) [fill = green!30, point, minimum width  = 0.15*\unit, minimum height = 0.15*\unit] (i1) {};
    \node at (4.5+2.6, 2.5) [fill = red!30, point, minimum width  = 0.15*\unit, minimum height = 0.15*\unit] (i2) {};
    
    \path[-stealth, -, line width = 0.8pt, bend left = 23.5] (i1) edge [] node {} (i2);
    \path[-stealth, -, line width = 0.8pt, bend right = 23.5] (i1) edge [] node {} (i2);
    \draw [-stealth, dashed, -, semithick] (4.5+2.6,0) -- (4.5+2.6,2.5);
    
    \node [left] at (i1) {$\cl_X(x, x')$};
    \node [above] at (i2) {$\cu_X(x, x')$};
    
    \node [above] at (4.5+1.25, 2.2) {$u_X^\CO(x, x')$};
    \node [below] at (4.5+1.85, 1) {$u_X^\CL(x, x')$};
    
    \node [below] at (4.5+0.15,0) {$0$};
    \node [below] at (4.5+2.6,0) {$1$};
    \node at (3.3, 1.5) {\huge $\Rightarrow$};
    
\end{tikzpicture}} }
\caption{Summary of admissible hierarchical clustering methods for metric spaces with distances given by intervals. Given a pair of nodes $x \neq x' \in X$, in metric space, single linkage clustering ultrametric $u_X^\SL(x, x')$ is the unique output satisfying (A1)-(A2). In general scenarios, there exists a family of admissible clustering methods (the blue region on the right). When the confidence level is at the extreme points ($\alpha \in \{0, 1\}$), singe linkage clustering ultrametric $\cu_X(x, x')$ and $\cl_X(x, x')$ are the respective unique method (red and green points). For a given confidence level $0 < \alpha < 1$, cluster-and-combine output $u_X^\CO(x, x')$ and combine-and-clustering output $u_X^\CL(x, x')$ provide bounds on all admissible methods.} 
\label{fig_summary}
\end{figure}

%
\begin{corollary}\label{coro_same}
Let $\ccalH: \ccalM \rightarrow \ccalU$ be a hierarchical clustering method. If $\ccalH$ satisfies axioms (A1) and (A2) then $\ccalH \equiv \ccalH^\SL$.
\end{corollary}

%
The uniqueness results claimed by Corollaries \ref{coro_alpha_0_1} and \ref{coro_same} can be considered as generalization of the uniqueness statement of single linkage clustering for metric space in \cite[Theorem 18]{carlsson2010}. When we take the most conservative belief and consider distance between points as their distance upper bounds $\du_X$, the only admissible method is the single linkage clustering applied onto the upper bounds $\du_X$. On the other hand, when we are the most liberate and trust the information conveyed in the distance lower bounds $\dl_X$, single linkage clustering applied onto $\dl_X$ is the unique admissible method. In metric space $(X, d_X)$ with $d_X:= \du_X = \dl_X$, irrespective of our belief of $\alpha$, the unique clustering method is the single linkage clustering applied onto $d_X$. Therefore, we can summarize the space of admissible hierarchical clustering in Fig.~\ref{fig_summary}. The unique clustering method $\ccalH^\SL$ in metric spaces becomes a space of admissible methods when distances are given by intervals. When the confidence interval is at the extreme points ($\alpha \in \{0 ,1\}$), the uniqueness of admissible methods is provided by Corollary \ref{coro_alpha_0_1}. For general confidence level $0 < \alpha < 1$, the admissible methods are not unique; cluster-and-combine as well as combine-and-clustering methods provide uniformly minimal and maximal bounds, which is established in Theorem \ref{thm_bounds}. 
We note that, given a specific confidence level $\alpha$, the output of the admissible methods do not differ much -- see examples in Section \ref{sec_application}.

%
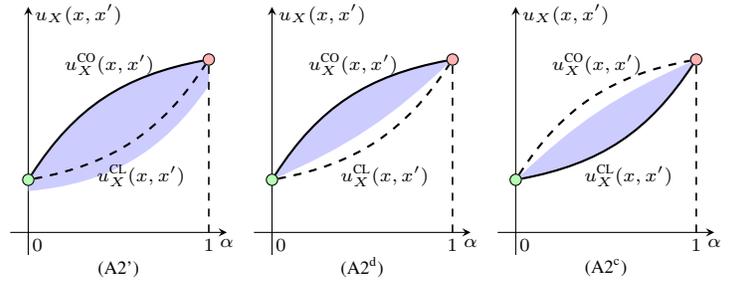
\begin{figure}[t]
\centerline{

\def \thisplotscale {1}

\def \unit {\thisplotscale cm}
\def \xsep {2.7}
\def \d {\text{d}}
\def \c {\text{c}}

\tikzstyle {vertex} = [circle, 
                       draw,
                       minimum width = 0.7*\unit,
                       minimum height = 0.7*\unit,
                       anchor=center,
                       font=\small]
\tikzstyle {light}  = [opacity = 0.1]
\tikzstyle{bigvertex} = [vertex, 
                        minimum width  = 0.7*\unit,
                        minimum height = 0.7*\unit]
                        
{\scriptsize \begin{tikzpicture}[x = 1.2*\unit, y = 1*\unit]

   
     \draw [-stealth] (-0.2,0) -- (2.2, 0) node [below, at end] {$\alpha$};
    \draw [-stealth] (0, -0.3) -- (0, 3);
    \node [right] at (0, 2.9) {$u_X(x, x')$};
    \node [below] at (1, -0.3) {(A2')};
    
    \path [fill = blue!20] (0, 0.55) to [bend right = 25] (2, 1.95) to (2, 2.3) to [bend right = 25] (0, 0.7); 
    \draw [-stealth, dashed, -, semithick] (2,0) -- (2,2.3);
    \node at (0, 0.7) [fill = green!30, point, minimum width  = 0.15*\unit, minimum height = 0.15*\unit] (i1) {};
    \node at (2, 2.3) [fill = red!30, point, minimum width  = 0.15*\unit, minimum height = 0.15*\unit] (i2) {}; 
    
    \path[-stealth, -, line width = 0.8pt, bend left = 23.5] (i1) edge [] node {} (i2);
    \path[-stealth, -, line width = 0.8pt, bend right = 23.5, dashed] (i1) edge [] node {} (i2);   
    
    \node [above] at (0.9, 2.0) {$u_X^\CO(x, x')$};
    \node [below] at (1.25, 1.0) {$u_X^\CL(x, x')$};
    
    \node [below] at (0.1,0) {$0$};
    \node [below] at (2,0) {$1$};
    
   
     \draw [-stealth] (\xsep-0.2,0) -- (\xsep+2.2, 0) node [below, at end] {$\alpha$};
    \draw [-stealth] (\xsep+0, -0.3) -- (\xsep+0, 3);
    \node [right] at (\xsep+0, 2.9) {$u_X(x, x')$};
    \node [below] at (\xsep+1, -0.23) {(A2$^\d$)};
    
    \path [fill = blue!20] (\xsep+0, 0.7) to [bend right = 10] (\xsep+2, 2.3) to [bend right = 25] (\xsep+0, 0.7); 
    \draw [-stealth, dashed, -, semithick] (\xsep+2,0) -- (\xsep+2,2.3);
    \node at (\xsep+0, 0.7) [fill = green!30, point, minimum width  = 0.15*\unit, minimum height = 0.15*\unit] (i1) {};
    \node at (\xsep+2, 2.3) [fill = red!30, point, minimum width  = 0.15*\unit, minimum height = 0.15*\unit] (i2) {}; 
    
    \path[-stealth, -, line width = 0.8pt, bend left = 23.5] (i1) edge [] node {} (i2);
    \path[-stealth, -, line width = 0.8pt, bend right = 23.5, dashed] (i1) edge [] node {} (i2);   
    
    \node [above] at (\xsep+0.9, 2.0) {$u_X^\CO(x, x')$};
    \node [below] at (\xsep+1.25, 1.0) {$u_X^\CL(x, x')$};
    
    \node [below] at (\xsep+0.1,0) {$0$};
    \node [below] at (\xsep+2,0) {$1$};
    
   
     \draw [-stealth] (2*\xsep-0.2,0) -- (2*\xsep+2.2, 0) node [below, at end] {$\alpha$};
    \draw [-stealth] (2*\xsep+0, -0.3) -- (2*\xsep+0, 3);
    \node [right] at (2*\xsep+0, 2.9) {$u_X(x, x')$};
    \node [below] at (2*\xsep+1, -0.26) {(A2$^\c$)};
    
    \path [fill = blue!20] (2*\xsep+0, 0.7) to [bend right = 25] (2*\xsep+2, 2.3) to [bend right = 10] (2*\xsep+0, 0.7); 
    \draw [-stealth, dashed, -, semithick] (2*\xsep+2,0) -- (2*\xsep+2,2.3);
    \node at (2*\xsep+0, 0.7) [fill = green!30, point, minimum width  = 0.15*\unit, minimum height = 0.15*\unit] (i1) {};
    \node at (2*\xsep+2, 2.3) [fill = red!30, point, minimum width  = 0.15*\unit, minimum height = 0.15*\unit] (i2) {}; 
    
    \path[-stealth, -, line width = 0.8pt, bend left = 23.5, dashed] (i1) edge [] node {} (i2);
    \path[-stealth, -, line width = 0.8pt, bend right = 23.5] (i1) edge [] node {} (i2);   
    
    \node [above] at (2*\xsep+0.9, 2.0) {$u_X^\CO(x, x')$};
    \node [below] at (2*\xsep+1.25, 1.0) {$u_X^\CL(x, x')$};
    
    \node [below] at (2*\xsep+0.1,0) {$0$};
    \node [below] at (2*\xsep+2,0) {$1$};

\end{tikzpicture}} }
\caption{Admissible hierarchical clustering methods given other possible construction of Axiom of Transformation discussed in Section \ref{sec_other_A2}. On the left, axiom (A2') is a weaker requirement compared to (A2) and therefore (A1)-(A2') yields a larger set of admissible clustering methods. In specific, combine-and-cluster output $u_X^\CO(x, x')$ is still a global maximum but $u_X^\CL(x, x')$ may no longer be a global minimum. In the middle and on the right, axioms ($\text{A2}^\d$) as well as ($\text{A2}^\c$) are a stronger requirements compared to (A2) and therefore their respective combination with (A1) gives a smaller set of admissible clustering methods.} 
\label{fig_other_A2}
\end{figure}

%
\subsection{Other Constructions of Axiom of Transformation}\label{sec_other_A2}

In Axiom of Transformation (A2), we require the output ultrametric to abide to $u_X(x, x') \geq u_Y(\phi(x), \phi(x'))$ when the map $\phi$ is $\alpha$-distance-reducing, i.e. satisfying \eqref{eqn_distance_reducing_d} and \eqref{eqn_distance_reducing_c}. Even though we justify that Eqs. \eqref{eqn_distance_reducing_d} and \eqref{eqn_distance_reducing_c} are equivalent to the natural condition on the map $\phi$ such that $d_X(x, x') \geq d_Y(\phi(x), \phi(x'))$ when restrict attention onto metric spaces $\ccalM$, some readers may find such requirement on $\ccalI$ is not highly intuitive and are curious to see what would work for other constructions of axiom of transformation. In this section, we consider other generalizations of axiom of transformation and the admissible clustering methods induced by them. We focus on presenting results and omit proofs.

We start by considering the following construction.
\begin{mylist}
\item [\it (A2') Axiom of Transformation (Alternative).] Consider $I_X = (X, \dl_X, \du_X)$ and $I_Y = (Y, \dl_Y, \du_Y)$ and a given map $\phi: X \rightarrow Y$ such that 
\begin{equation}\begin{aligned}\label{eqn_alternative_axiom_of_transofmration}
    \dl_X(x, \!x') \!\geq\! \dl_Y(\phi(x), \!\phi(x')), \du_X(x, \!x') \!\geq\! \du_Y(\phi(x), \!\phi(x')),
\end{aligned}\end{equation}
for any nodes $x \neq x'$. The ultrametrics $u_X = \ccalH(X, \dl_X, \du_X)$ and $u_Y = \ccalH(Y, \dl_Y, \du_Y)$ are said to abide to the axiom of transformation (alternative) if $u_X(x, x') \geq u_Y(\phi(x), \phi(x'))$. 
\end{mylist}

Note that the requirement on the map \eqref{eqn_alternative_axiom_of_transofmration} would imply $\phi$ is a $\alpha$-distance-reducing map. Hence, because compared to (A2), (A2') implies the same output $u_X(x, x') \geq u_Y(\phi(x), \phi(x'))$ under a stricter requirement on $\phi$, (A2') is a weaker condition than (A2) and therefore the admissible clustering methods satisfying (A1)-(A2') would be richer. In specific, as illustrated on the left of Fig.~\ref{fig_other_A2}, combine-and-cluster output $u_X^\CO(x, x')$ and cluster-and-combine output $u_X^\CL(x, x')$ are still admissible; $u_X^\CO(x, x')$ is a global maximum but we could not verify that $u_X^\CL(x, x')$ is still a global minimum. There might be other admissible methods yielding output $u_X(x, x')$ which is smaller than $u_X^\CL(x, x')$.

In (A2), we say a map is $\alpha$-distance-reducing if it satisfies both \eqref{eqn_distance_reducing_d} and \eqref{eqn_distance_reducing_c}. Investigate the construction for maps that satisfy only a single requirement of them yields the two possible ways to construct different axioms of transformation as we state next.
\begin{mylist}
\item [\it ($\text{A2}^\d$) Axiom of Transformation (Distance).] Consider $I_X$, $I_Y $, $0 \leq \alpha \leq 1$, and a map $\phi: X \rightarrow Y$ such that 
\begin{equation}\begin{aligned}\label{eqn_axiom_of_transofmration_distance}
    \hhatd_X(x, x') \geq \hhatd_Y(\phi(x), \phi(x'))
\end{aligned}\end{equation}
for any $x \neq x'$. The outputs $u_X$ and $u_Y$ are said to satisfy the axiom of transformation (distance) if $u_X(x, x') \geq u_Y(\phi(x), \phi(x'))$. 
\item [\it ($\text{A2}^\c$) Axiom of Transformation (Chain Costs).] Consider $I_X$, $I_Y $, $0 \leq \alpha \leq 1$, and a map $\phi: X \rightarrow Y$ such that 
\begin{equation}\begin{aligned}\label{eqn_axiom_of_transofmration_chain}
    \hhatc_X(x, x') \geq \hhatc_Y(\phi(x), \phi(x')), \forall x \neq x' \in X
\end{aligned}\end{equation}
The outputs $u_X$ and $u_Y$ are said to satisfy the axiom of transformation (chain costs) if $u_X(x, x') \geq u_Y(\phi(x), \phi(x'))$. 
\end{mylist}

For ($\text{A2}^\c$), even though the requirement in \eqref{eqn_axiom_of_transofmration_chain} is with respect to the combination of minimum chain costs and is different from distance bounds as in \eqref{eqn_axiom_of_transofmration_distance}, it can be shown that single linkage clustering $\ccalH^\SL$ is still the only admissible method satisfying  (A1)-($\text{A2}^\c$) when we restrict attention onto metric spaces.

Compared to (A2), both ($\text{A2}^\d$) and ($\text{A2}^\c$) induce the same output $u_X(x, x') \geq u_Y(\phi(x), \phi(x'))$ under weaker requirements on $\phi$. Consequently, both ($\text{A2}^\d$) and ($\text{A2}^\c$) are more stringent conditions than (A2). This implies that, compared to the the admissible clustering methods satisfying (A1)-(A2), the admissible methods satisfying (A1)-($\text{A2}^\d$) as well as methods satisfying (A1)-($\text{A2}^\c$) would be smaller. Indeed, as illustrated in the middle of Fig.~\ref{fig_other_A2}, for the axioms (A1)-($\text{A2}^\d$), combine-and-cluster output $u_X^\CO(x, x')$ is admissible but cluster-and-combine output $u_X^\CL(x, x')$ is not. In analogy, as on the right of Fig.~\ref{fig_other_A2}, for the axioms (A1)-($\text{A2}^\c$), $u_X^\CL(x, x')$ is admissible but $u_X^\CO(x, x')$ is not.  

We focus our analysis on axioms (A1)-(A2) because we believe $u_X^\CL(x, x')$ and $u_X^\CO(x, x')$ are reasonable clustering methods in metric spaces with distance given by intervals and should be included. Besides, we would like to have a statement in Theorem \ref{thm_value_implies_minimum_separation} for minimum separation and an observation as provided in Corollary \ref{coro_alpha_0_1} that when $\alpha \in \{0, 1\}$, the admissible methods would be unique given by the single linkage clustering methods applied onto the distance upper or lower bounds. 

%
\section{Applications}\label{sec_application}

We illustrate the usefulness of clustering theory developed in previous sections through numerical experiments in both synthetic scenario (Section \ref{sec_synthetic}) and real world dataset (Section \ref{sec_real_world}).

%
\begin{figure*}[t]
	\centering		
	\begin{minipage}[h]{0.32\linewidth}
		\centering
		\includegraphics[trim=0.6cm 0.2cm 1.5cm 0.5cm, clip=true, width=1 \textwidth]{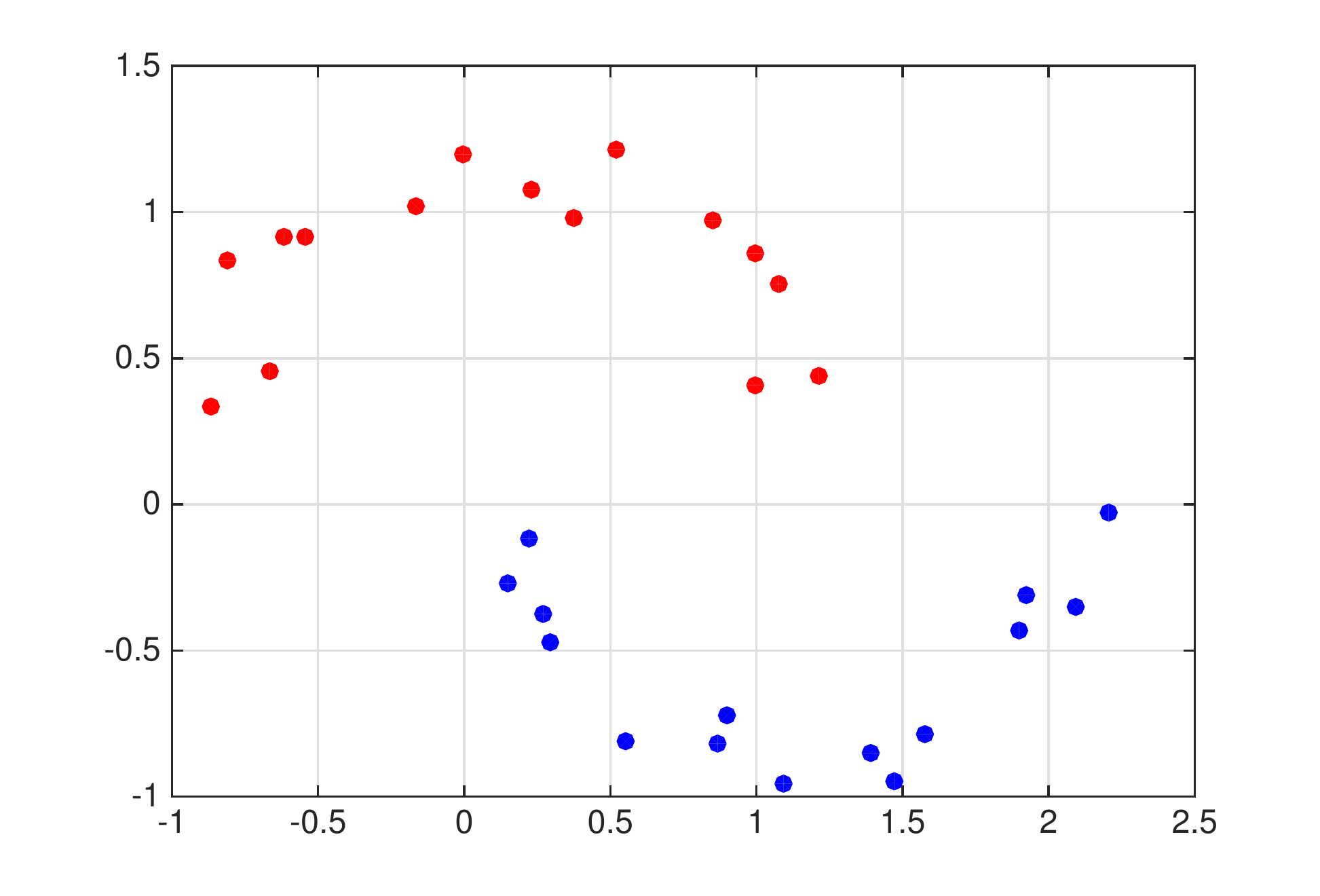}
		\scriptsize (a)
	\end{minipage}
	\begin{minipage}[h]{0.32\linewidth}
		\centering
		\includegraphics[trim=0.6cm 0.2cm 1.5cm 0.5cm, clip=true, width=1 \textwidth]{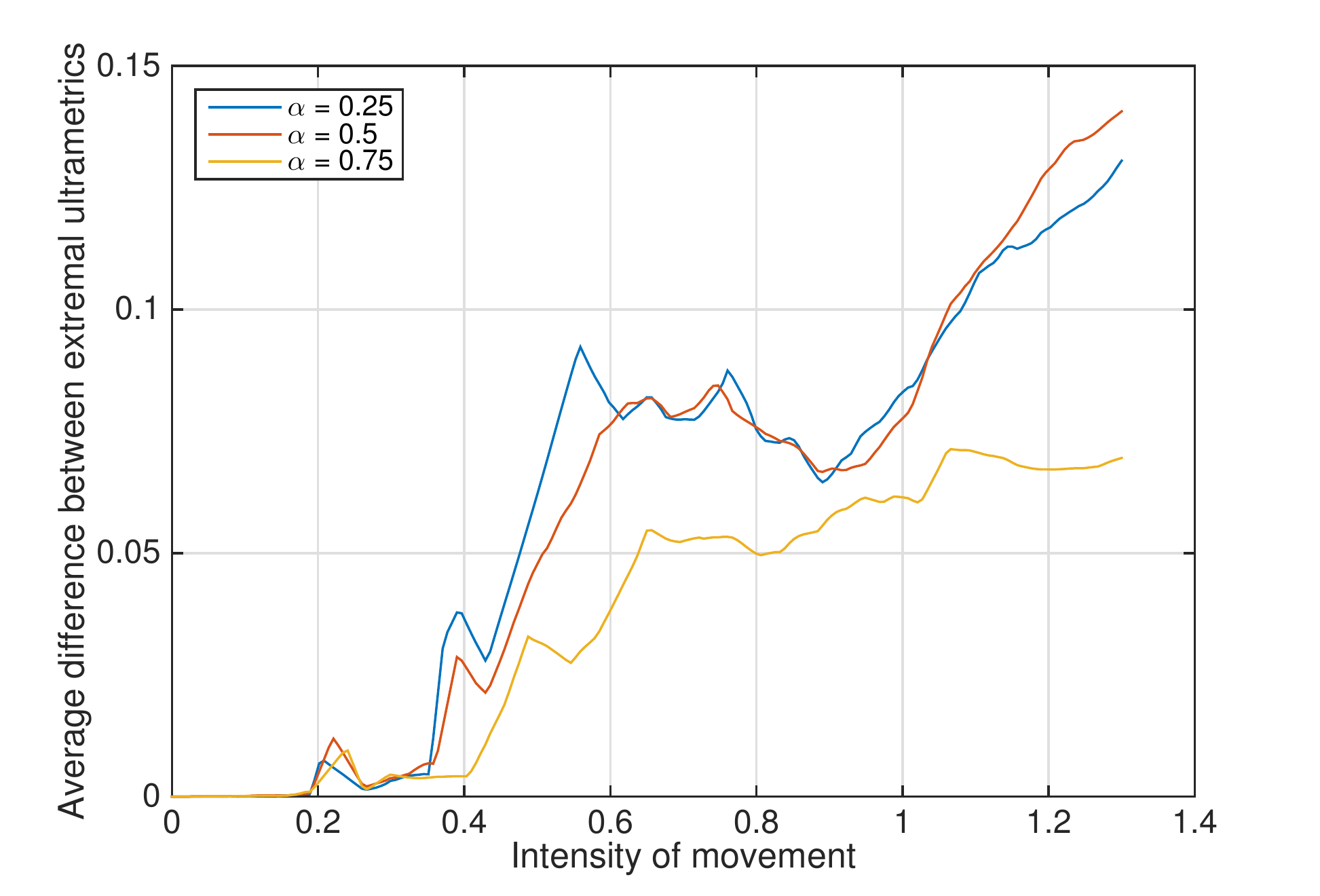}
		\scriptsize (b)
	\end{minipage}
	\begin{minipage}[h]{0.32\linewidth}
		\centering
		\includegraphics[trim=0.6cm 0.2cm 1.5cm 0.5cm, clip=true, width=1 \textwidth]{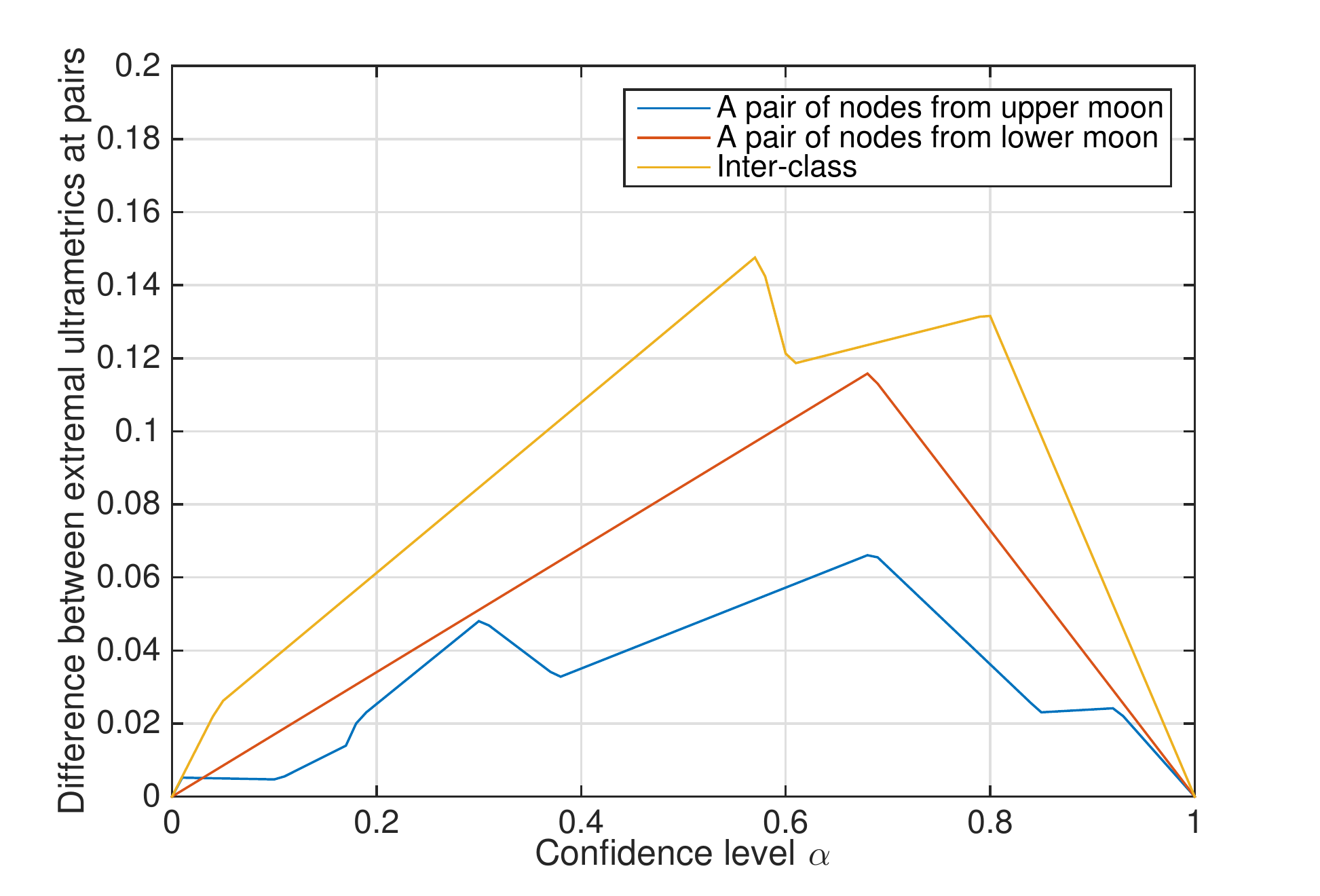}
		\scriptsize (c)
	\end{minipage}
	\caption{Synthetic experiment of clustering of moving points based on multiple snapshots. (a) Initial positions of points, which correspond to two half moons. (b) Relationship between the average difference $u_X^\CO(x, x') - u_X^\CL(x, x')$ of the two extremal clustering methods across all pairs of nodes $x \neq x' \in X$ and the intensity of movement $\sigma^2$. The difference in $u_X^\CO(x, x') - u_X^\CL(x, x')$ increases but is significantly smaller comparing to $\du_X(x, x') - \dl_X(x, x')$. (c) Relationship between the difference $u_X^\CO(x, x') - u_X^\CL(x, x')$ at three different pairs of points and the confidence level $\alpha$.}
	\label{fig_synthetic}
\end{figure*}

%
\begin{figure*}[t]
	\centering		
	\begin{minipage}[h]{0.32\linewidth}
		\centering
		\includegraphics[trim=1cm 0.5cm 1.8cm 0.5cm, clip=true, width=1 \textwidth]{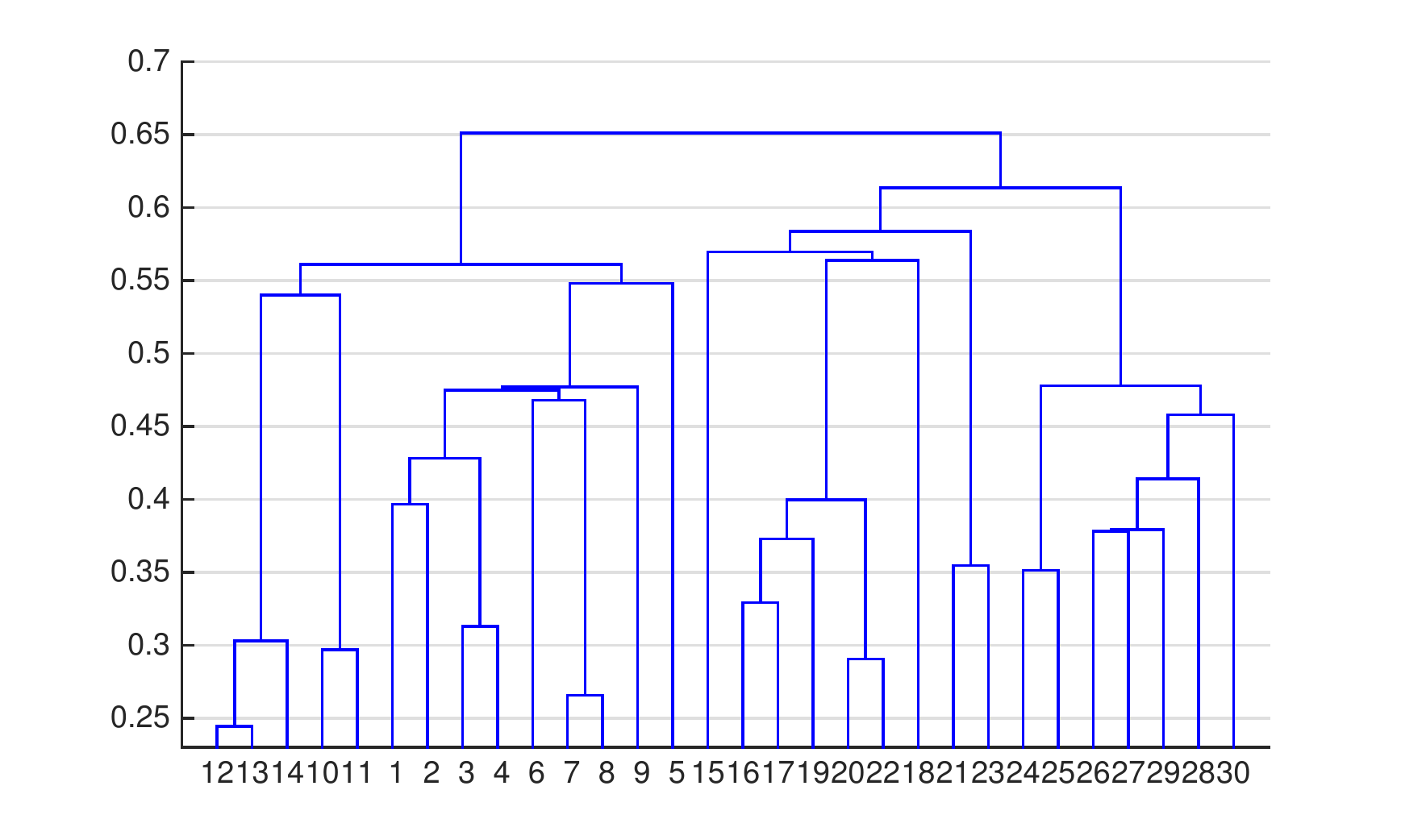}
		\scriptsize (a) Cluster-and-combine dendrogram
	\end{minipage}
	\begin{minipage}[h]{0.32\linewidth}
		\centering
		\includegraphics[trim=1cm 0.5cm 1.8cm 0.5cm, clip=true, width=1 \textwidth]{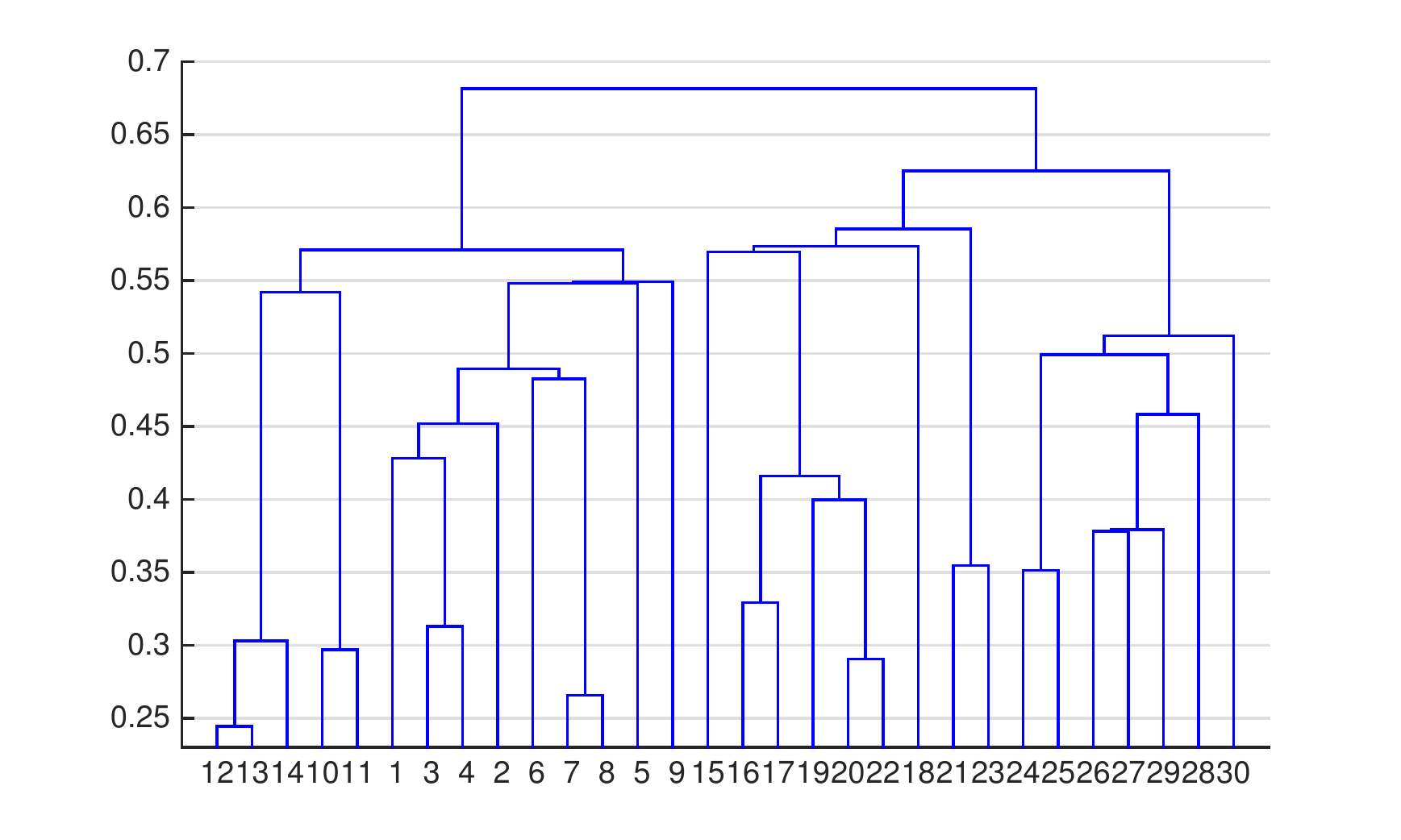}
		\scriptsize (b) Combine-and-cluster dendrogram
	\end{minipage}
	\begin{minipage}[h]{0.32\linewidth}
		\centering
		\includegraphics[trim=1cm 0.5cm 1.8cm 0.5cm, clip=true, width=1 \textwidth]{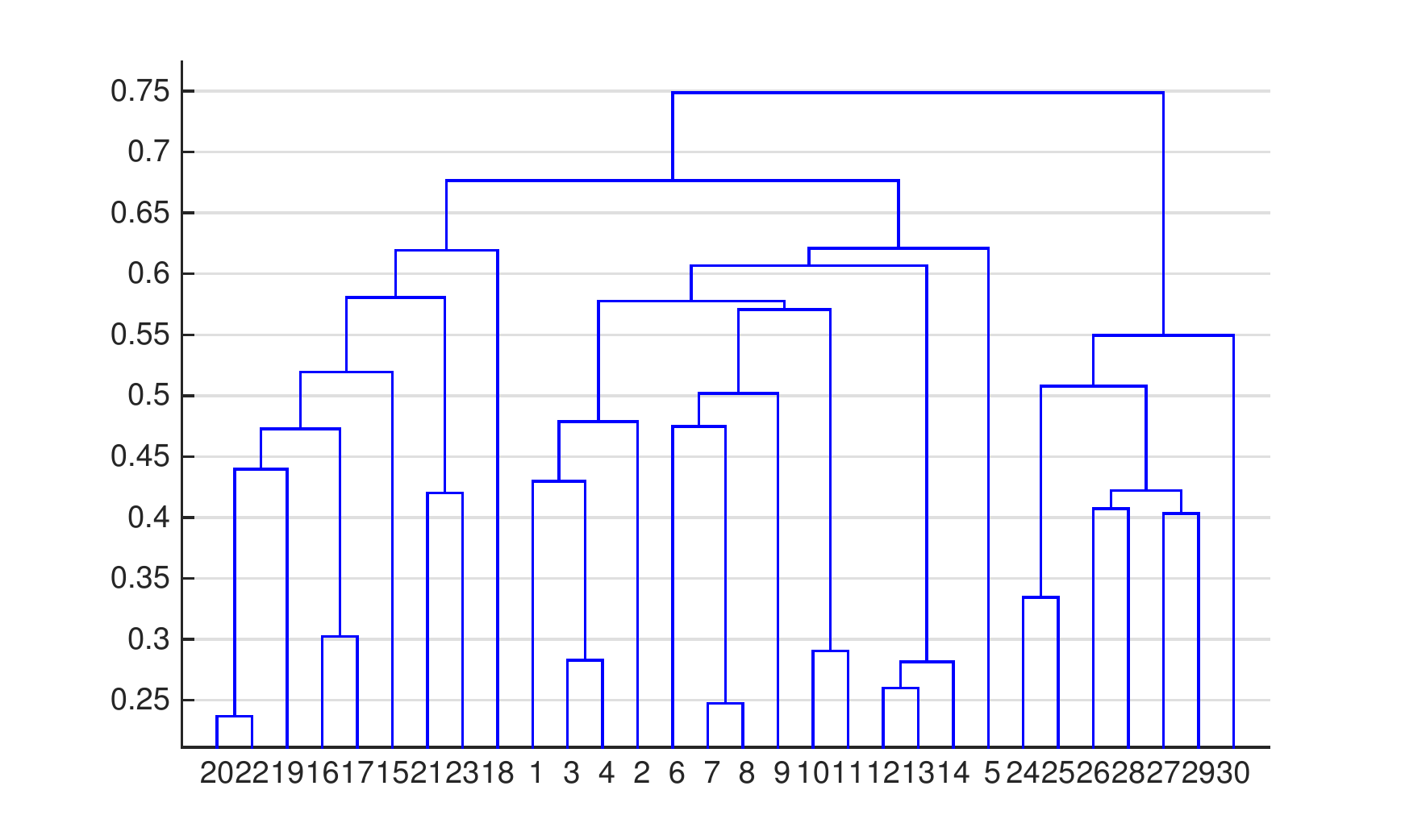}
		\scriptsize (c) Benchmark: dendrogram via mean distance
	\end{minipage}
	\caption{Resulting dendrograms of (a) cluster-and-combine method and (b) combine-and-cluster method applied upon the triplet $(X, \dl_X, \du_X)$ in the synthetic experiment. (c) Benchmark dendrogram: single linkage applied upon the mean distance between any pair of nodes $\tdd_X(x, x') = \frac{1}{T}\sum_{t = 1}^T d_X^t(x, x')$.}
	\label{fig_synthetic_dendrogram}
\end{figure*}

%
\subsection{Clustering of moving points by snapshots}\label{sec_synthetic}
	
We consider the clustering of $n$ moving points in a two-dimensional plane with the initial coordinate of the $i$-th point represented by $\bbp_i^0 \in \reals^2$. Points are moving in the plane and we have $T$ snapshots with $\bbp_i^t \in \reals^2$ denoting the coordinate of the $i$-th point at the $t$-th snapshot. We assume that the directions of movement of points are completely random and therefore model the observation as $\bbp_i^t := \bbp_i^{t-1} + \bbepsilon$ for any $i$ and any time point $1 \leq t \leq T$, where $\bbepsilon \in \reals^{2}$ is a two-dimensional independent zero-mean Gaussian random variable with covariance matrix $\sigma^2 \bbI$. Having no knowledge about the starting coordinates, we would like to evaluate clustering based on observations $\{ \bbp_i^t \}_{i = 1, \dots, n, t = 1, \dots, T}$. To do so, we consider the node set $X$ where $x_i \in X$ denotes the $i$-th point $\bbp_i$, and use $d_X^t(x_i, x_j) = \|\bbp_i^t - \bbp_j^t\|_2$ to represent the distance between the $i$-th and the $j$-th points at the $t$-th snapshot. Then we define metric space with distances given by intervals $(X, \dl_X, \du_X)$ such that given a pair of nodes $x_i \neq x_j$, we set the distance lower bound $\dl_X(x_i, x_j) = \min_{1 \leq t \leq T} d_X^t(x_i, x_j)$ as the minimum distance between the pair at all snapshots. Similarly, we define the distance upper bound $\du_X(x_i, x_j) = \max_{1 \leq t \leq T} d_X^t(x_i, x_j)$ as the maximum distance between the pair among all snapshots. Clustering methods are then applied upon the triplet $(X, \dl_X, \du_X)$. 

As an example, we consider $n = 30$ points whose initial coordinates form two half moons (Fig.~\ref{fig_synthetic} (a)), and investigate $T = 10$ snapshots of these moving points. We apply cluster-and-combine clustering $\ccalH^\CL$ and combine-and-cluster clustering $\ccalH^\CO$ onto the distance bounds $(X, \dl_X, \du_X)$. The average difference between the output ultrametrics $u_X^\CO(x, x') - u_X^\CL(x,x')$ across all pairs of nodes $x \neq x' \in X$ with respect to the intensity of movement, i.e. the variance $\sigma^2$ of $\bbepsilon$, is displayed in Fig.~\ref{fig_synthetic} (b) at three different $\alpha$. The average difference generally increases with the intensity, however, does not increase significantly. As a comparison, the average difference between the distance bounds $\du_X(x, x') - \dl_X(x, x')$ is $0.6353$; even at relatively intense movement with $\sigma^2 \geq 1.2$, the average difference between the ultrametrics $u_X^\CO(x, x') - u_X^\CL(x, x')$ is less than $20\%$ of the difference between the input distance bounds $\du_X(x, x') - \dl_X(x, x')$. The relationship between the difference $u_X^\CO(x, x') - u_X^\CL(x, x')$ at three different pairs of points and $\alpha$ is plotted in Fig.~\ref{fig_synthetic} (c) where $\sigma^2$ is set to $0.9$. At the extremal confidence level with $\alpha \in \{0, 1\}$, ultrametrics $u_X^\CO(x, x')$ and $u_X^\CL(x, x')$ coincide, verifying the result in Corollary \ref{coro_alpha_0_1}. Besides, the difference $u_X^\CO(x, x') - u_X^\CL(x, x')$ is not very high. Combining this with Theorem \ref{thm_bounds}, the outputs of all feasible hierarchical clustering methods do not differ by much at a given $\alpha$. 

Finally, Fig.~\ref{fig_synthetic_dendrogram} (a) and (b) show the output dendrograms of cluster-and-combine and combine-and-cluster methods, respectively. The variance parameter $\sigma^2$ of movement $\bbepsilon$ is set as $0.4$ and the confidence level $\alpha$ as $0.5$. Nodes $1$ to $15$ correspond to points in the upper moon regarding their initial coordinates; nodes $16$ to $30$ correspond to points in the lower moon. It can be seen from Fig.~\ref{fig_synthetic} (a) and (b) that (i) both $\ccalH^\CL$ and $\ccalH^\CO$ yield the desired output (two clusters correspond to two half moons), with only one point ($15$) gets misclassified, (ii) limited difference exists between the two dendrograms, and (iii) points closer in their initial positions (e.g. points $1$ to $9$, $10$ to $14$, and $24$ to $30$) tend to be clustered together at lower resolutions. As a benchmark, we consider the mean distance between any pair of nodes $\tdd_X(x, x') = \frac{1}{T}\sum_{t = 1}^T d_X^t(x, x')$ and apply single linkage upon $(X, \tdd_X)$. Fig.~\ref{fig_synthetic_dendrogram} (c) shows the resulting dendrogram, which fails to identify the clusters correctly.

%
\subsection{Clustering of networks via distance bounds}\label{sec_real_world}

 In this section, we apply the clustering methods to bounds on metrics in the space of networks. The problem of comparing and clustering networks is interesting on its own \cite{ali2014, choobdar12}. In our previous contributions \cite{Huang15, Huang15a}, we have defined network differences derived from an underlying metric in the space of networks. We briefly review notations here. We consider network in the form of $N_Z = (Z, r_Z)$ with $Z$ denotes the set of points in the network and $r_Z : Z \times Z \rightarrow \reals^+$ a relationship function. For points $z$ and $z'$, the value $r_Z(z, z')$ is intended to represent similarity between the pair. The function $r_Z$ is nonnegative and symmetric, satisfies $r_Z(z, z') = 0$ if and only if $z = z'$, however, does not necessarily satisfy the triangle inequality. The set of all networks is denoted as $\ccalN$. Two networks $N_Z$ and $N_W$ are said isomorphic if there exists a bijection $\pi: Z \rightarrow W$ such that for all $x_{0:k} \in X^{k+1}$ we have $r_W (\pi(z), \pi(z')) = r_Z(z, z')$ for any $z \neq z' \in Z$. Since the map $\pi$ is bijective, isomophism can only be satisfied when $Z$ is a permutation of $W$. When networks $N_Z$ and $N_W$ are isomorphic we write $N_Z \cong N_W$. The space of networks where isomorphic networks are represented by the same element is termed the set of networks modulo isomorphism and denoted by $\ccalN \mod \cong$. The space $\ccalN \mod \cong_k$ can be endowed with a valid metric \cite{Huang15}. The definition of this distance requires introducing the notion of correspondence \cite[Def. 7.3.17]{Burago01}:

%
\begin{definition}\label{dfn_correspondence}
A correspondence between two sets $X$ and $Y$ is a subset $C \subseteq X \times Y$ such that $\forall~x \in X$, there exists $y \in Y$ such that $(x,y) \in C$ and $\forall~ y \in Y$ there exists $x \in X$ such that $(x,y) \in C$. The set of all correspondences is denoted as $\ccalC(X,Y)$.
\end{definition}

%
A correspondence in Definition \ref{dfn_correspondence} connects node sets $X$ and $Y$ so that every element of one set has at least one correspondent in the other. We now define the distance between two networks by selecting the correspondence that makes them the most similar.

%
\begin{definition}\label{dfn_d_N}
Given networks $N_Z$ and $N_W$ and a correspondence $C$ between $Z$ and $W$, the network difference with respect to $C$ is 
\begin{align}\label{eqn_d_N_prelim} 
   \Gamma_{Z,W} (C)
      := \max_{(z, w), (z', w') \in C}  \
            \left| r_Z(z, z') - r_W (w, w') \right|,
\end{align}
The network distance between $N_Z$ and $N_W$ is then defined as
\begin{align}\label{eqn_d_N}
   d_\ccalN (N_Z, N_W) := \min_{C \in \ccalC(Z,W)} \
   \left\{ \Gamma_{Z,W} (C) \right\}. \end{align}\end{definition}

%
For a given correspondence $C \in \ccalC(Z,W)$ the network difference $\Gamma_{Z,W}(C)$ selects the maximum distance difference $|r_Z(z, z') - r_W (w, w')|$ among all pairs of correspondents. The distance in \eqref{eqn_d_N} is defined by selecting the correspondence that minimizes these maximal differences. 
 The metric distances defined here have been applied to compare networks with small number of nodes and have succeeded in identifying collaboration patterns of coauthorship networks \cite{Huang15, Huang15b}. However, because they have to consider all possible node correspondences, network distances are difficult to compute when the number of nodes in the networks is large. To resolve such problem, we mapped networks to filtrations of simplicial complexes and demonstrated that the difference between the homological features of their respective filtration can be used as a lower bound of $d_\ccalN$ \cite{Huang16, Huang16a}. Computational of homological features is fast \cite{Mischaikow13} and we have applied these lower bounds in comparing the coauthorship networks of engineering and mathematics journals. 

On the other hand, $\Gamma_{Z,W} (C)$ in \eqref{eqn_d_N_prelim} for any correspondence $C$ witnesses an upper bound on the distance $d_\ccalN$. Therefore, given a set of networks $X$ where the $i$-th element $x_i$ denotes a network $N_i$, we can evaluate the upper and lower bounds of network distance $d_\ccalN(N_i, N_j)$ for any pair of networks $N_i$ and $N_j$ in $X$ to yield a metric in the space of networks where distances are given by intervals $(X, \dl_X, \du_X)$. Clustering methods examined in the paper can then be applied towards the triplet to categorize networks.

As an example of network classification, we consider real-world brain networks of patients diagnosed with Fronto-Temporal Dementia (FTD) or Alzheimer's disease (AD) as well as healthy controls \cite{binnewijzend2014, medaglia2016}. For each network $N_Z = (Z, r_Z)$, $Z$ denotes the set of $119$ brain regions, where each region $z$ represents brain areas that are anatomically close and functionally similar. The relationship function $r_Z(z, z')$ denotes the number of connecting neuron streamlines between brain regions $z$ and $z'$. There is an underlying labeling of brain regions, and therefore given a pair of networks $N_Z = (Z, r_Z)$ and $N_\tdZ = (Z, r_\tdZ)$, it is reasonable to utilize the sum of differences on all connections 
\begin{align}\label{eqn_real_benchmark}
    d_b(N_Z, N_\tdZ) := \sum_{z \neq z'} |r_Z(z, z') - r_\tdZ(z, z')|,
\end{align}
as the difference between the pair; in fact, we use this as a benchmark with the methods proposed in the paper. Nonetheless, using the same labeling assumes that same brain region has exactly identical functionality across all subjects, while in reality brain region $z$ of one subject might be the most similar with brain region $z' \neq z$ of another subject. Motivated by this, we consider brain networks as unlabeled entities and evaluate the upper and lower bounds of network distance $d_\ccalN(N_Z, N_\tdZ)$ defined in Definition \ref{dfn_d_N}. In specific, the lower bound $\dl_X(N_Z, N_\tdZ)$ is established via the difference in the respective homological features. The upper bound $\du_X(N_Z, N_\tdZ)$ is constructed using the peculiar correspondence $C$
\begin{align}\label{eqn_real_upper_bound}
    \du_X (N_Z, N_\tdZ) := \max_{z \neq z'} |r_Z(z, z') - r_\tdZ(z, z')|.
\end{align}

\begin{figure*}[t]
	\centering		
	\begin{minipage}[h]{0.49\linewidth}
		\centering
		\includegraphics[trim=3.5cm 0.8cm 3.5cm 0.3cm, clip=true, width=1 \textwidth]{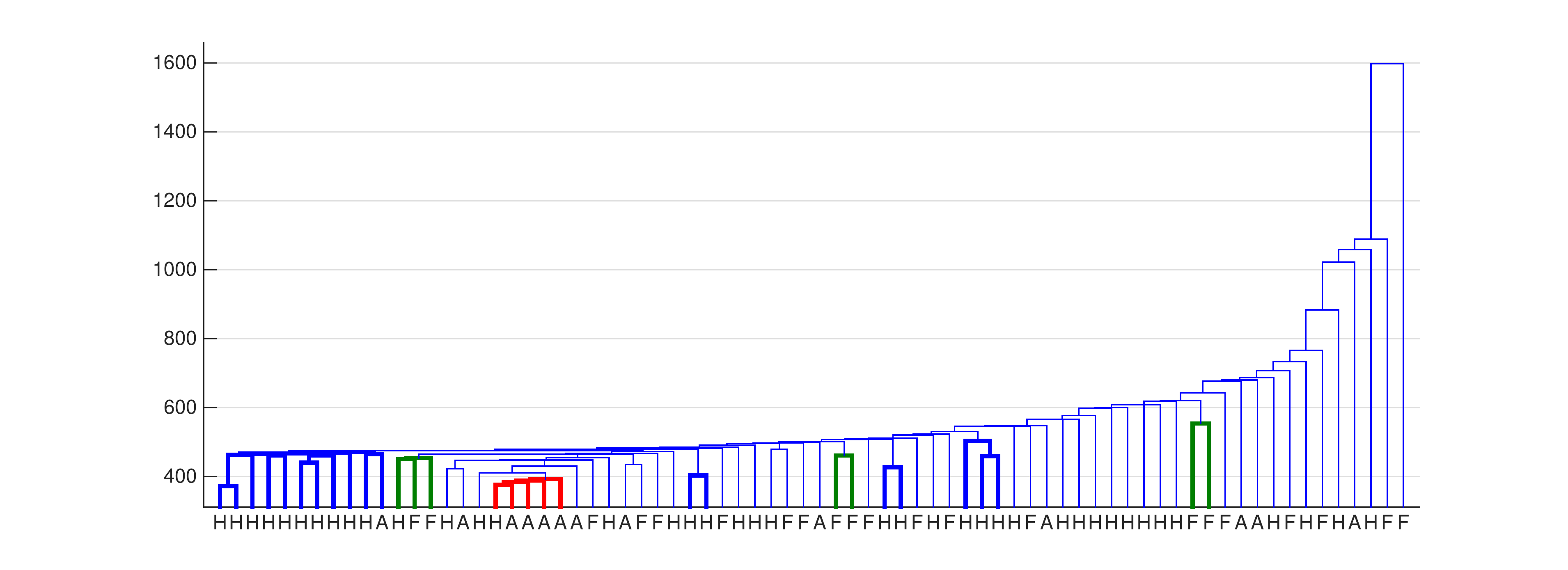}
		\scriptsize (a) Cluster-and-combine dendrogram
	\end{minipage}
	\begin{minipage}[h]{0.49\linewidth}
		\centering
		\includegraphics[trim=3.5cm 0.8cm 3.5cm 0.3cm, clip=true, width=1 \textwidth]{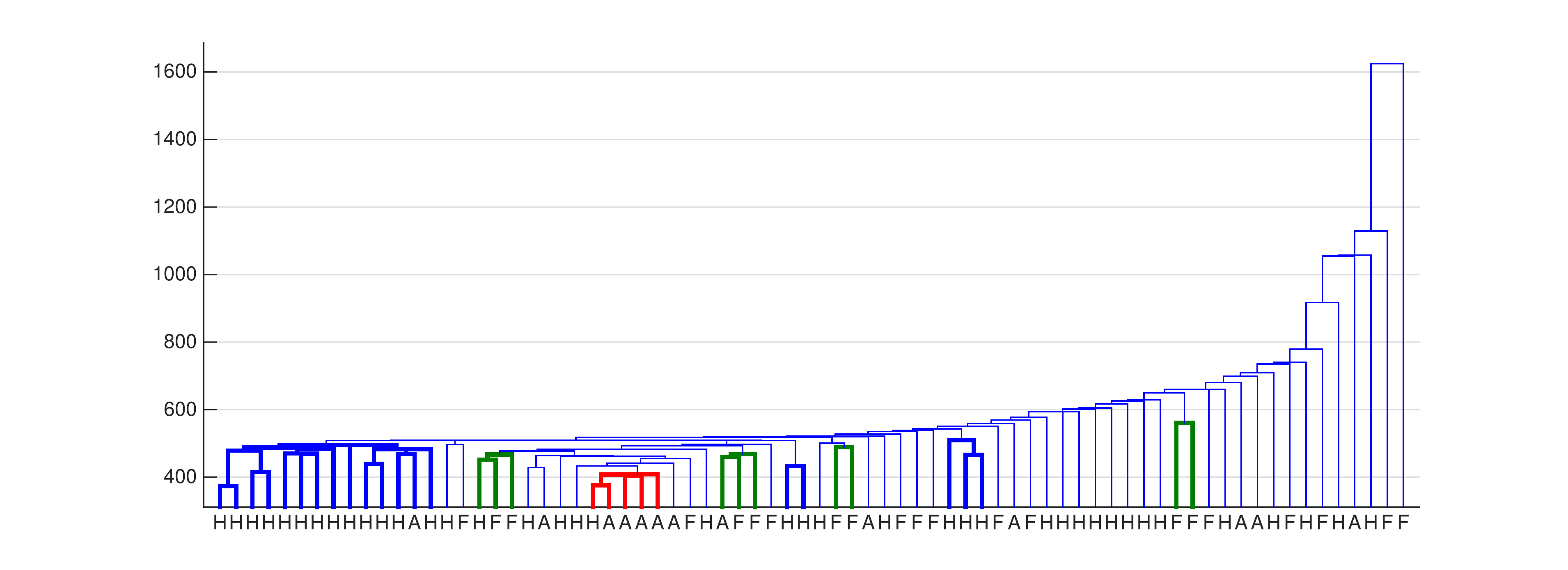}
		\scriptsize (b) Combine-and-cluster dendrogram
	\end{minipage}
	\caption{Resulting dendrograms of cluster-and-combine method (a) and combine-and-cluster method (b) applied upon the triplet $(X, \dl_X, \du_X)$ in the brain network application. Each node in the dendrograms represents the brain structural connectivity network of a participant, where healthy controls are labeled as `H', patients with FTD as `F', and patients with AD as `A'. The difference between the output dendrograms of the two methods is small. Besides, brain networks corresponding to patients with same behavior status tend to be clustered together at lower resolutions.}
	\label{fig_real_dendrogram}
\end{figure*}

We apply cluster-and-combine as well as combine-and-cluster methods upon the constructed metric in the space of networks with distances given by intervals $(X, \dl_X, \du_X)$ with confidence level $\alpha = 0.5$. Fig.~\ref{fig_real_dendrogram} shows the respective dendrogram, where healthy controls are labeled as `H', patients with FTD as `F', and patients with AD as `A'. Similar as in synthetic experiments, the difference between the resulting dendrogram of $\ccalH^\CL$ and $\ccalH^\CO$ is small. This observation combined with the guarantee established by Theorem \ref{thm_bounds} further demonstrates that the outputs of all feasible hierarchical clustering methods do not differ by much. Besides, networks corresponding to patients with same health status (`H' or `F' or `A') tend to be clustered together at lower resolution, which is highlighted by bold colored lines in the dendrograms. Finally, as a quantitative benchmark, we evaluate the error of unsupervised classification based on the output ultrametrics $u_X^\CL$ and $u_X^\CO$ in classifying healthy controls from patients with either FTD or AD, and compare its error with the similar investigation based on single linkage applied towards the benchmark difference $d_b(N_Z, N_\tdZ)$ defined in \eqref{eqn_real_benchmark}. The clustering methods $\ccalH^\CL$ and $\ccalH^\CO$ yield $28.38\%$ unsupervised classification error, which is slightly better than the $29.73\%$ error of the benchmark.

%

%
\section{Conclusions}\label{sec_conclusion}
We have developed a theory for hierarchically clustering metric spaces where distances are given by intervals. We begin by identifying simple and reasonable axioms of value 
and transformation
; we then proceeded to characterize the methods that abide to these properties. Two admissible methods were constructed and were proved to serve as upper and lower bounds in the space of all admissible methods. In metric spaces, the two methods were shown to boil down to the known fact that single linkage clustering is the unique admissible method; the space of admissible methods was illustrated to become richer when distances are given in intervals. We explored the practical usefulness by clustering moving nodes using multiple snapshots and by clustering networks denoting human brain structural connectivity via the upper and lower bounds of the network distances. The proposed methods succeeded in identifying the underlying clustering structures of moving points, and made a moderate success in associating brain connectivity networks to their neruroscitific behaviors.

\appendices
%
\section{Proof of Theorem \ref{thm_value_implies_minimum_separation}}\label{apx_proof_thm_2}

To prove Theorem \ref{thm_value_implies_minimum_separation}, for any $(X, \dl_X, \du_X)$ and $0 \leq \alpha \leq 1$, we would like to define a two-node space $(\{p, q\}, \du, \dl)$. Moreover, given a pair of distinct nodes $x \neq x' \in X$, we would like to find a $\alpha$-distance-reducing map $\phi_{x, x'} : X \rightarrow \{p, q\}$ from $X$ to the space of two nodes. In order to achieve that, denote $\dotx$ and $\dotx'$ as the pair achieving $\sep^\alpha(X, \dl_X, \du_X)$. For this pair of nodes, define
\begin{align}\label{eqn_proof_thm_minimum_separation_two_node_space_etal}
    \etal &:= \min_{C(\dotx, \dotx')} \max_{i \mid x_i \in C(\dotx, \dotx')} \dl_X (x_i, x_{i+1}),
\end{align}
and similarly for $\etau$ such that $\alpha \etau + (1 - \alpha) \etal$ is the same as $\sep^\alpha(X, \dl_X, \du_X)$. To construct a $\alpha$-distance-reducing map $\phixx$, first define $\phixx(x) = p$ and $\phixx(x') = q$; then their $\alpha$-separation in the respective node space satisfies
\begin{align}\label{eqn_proof_thm_minimum_separation_d_pq}
    s^\alpha_X(x, x') & \geq s^\alpha_X(\dotx, \dotx') =: s^\alpha_{p,q}(\phixx(x), \phixx(x')).
\end{align}
Now, for $\tdx$ different from $\dotx$ and $\dotx'$, $\phixx(\tdx)$ can take $p$ or $q$ arbitrarily. To see why this is valid, consider $\tdx \neq \tdx'$, where at least one of the nodes is neither $x$ nor $x'$, if $\phixx(\tdx) = \phixx(\tdx')$, then $\phixx$ is $\alpha$-distance-reducing. This follows because $s^\alpha_X(\tdx, \tdx') \geq 0 =: s^\alpha_{p,q}(\phixx(\tdx), \phixx(\tdx'))$, which is \eqref{eqn_distance_reducing_c}. Moreover,
\begin{align}\label{eqn_proof_thm_minimum_separatio_preparation}
    \alpha \du_X(\tdx, \tdx') + (1 - \alpha) \dl_X(\tdx, \tdx') &\geq s^\alpha_X(\tdx, \tdx'), \\
        \label{eqn_proof_thm_minimum_separatio_preparation_2}
            \dl_{p,q}(\phixx(\tdx), \phixx(\tdx')) = \du_{p,q}(\phixx(\tdx), &\phixx(\tdx')) = 0,
\end{align}
would yield us $\hhatd_X(\tdx, \tdx') \geq 0 =: \hhatd_{p,q}(\phixx(\tdx), \phixx(\tdx'))$, the condition in \eqref{eqn_distance_reducing_d}. On the other hand, if $\phixx(\tdx) \neq \phixx(\tdx')$,
\begin{equation}\begin{aligned}\label{eqn_proof_thm_minimum_separation_2}
    s^\alpha_X(\tdx, \tdx') & \geq s^\alpha_X(\dotx, \dotx') =: s^\alpha_X(\phixx(\tdx), \phixx(\tdx')),
\end{aligned}\end{equation}
which follows from the definition of $\dotx$ and $\dotx'$ as well as the construction of the two-node space $\Delta_2(\etau, \etal)$. This is the requirement in \eqref{eqn_distance_reducing_c}. Besides, notice
\begin{align}\label{eqn_proof_thm_minimum_separatio_preparation_3}
    \alpha \du + (1 - \alpha) \dl = s_{p,q}^\alpha(\phixx(\tdx), \phixx(\tdx')).
\end{align}
Combining \eqref{eqn_proof_thm_minimum_separatio_preparation}, \eqref{eqn_proof_thm_minimum_separation_2} and \eqref{eqn_proof_thm_minimum_separatio_preparation_3} yields the condition in \eqref{eqn_distance_reducing_d}, 
\begin{align}\label{eqn_proof_thm_minimum_separatio_preparation_3_d}
    \du_X(\tdx, \tdx') + (1 - \alpha) \dl_X(\tdx, \tdx') \geq \alpha \du + (1 - \alpha) \dl.
\end{align}
This shows that $\phi_{x, x'}$ is always a $\alpha$-distance-reducing map.

Denote $(\{p, q\}, u_{p, q}) = \ccalH(\Delta_2(\etau, \etal))$ as the ultrametric space obtained when apply the $\ccalH$ to the two-node space $\Delta_2(\etau, \etal)$. Since $\ccalH$ satisfies the Axiom of Value (A1), we must have 
\begin{equation}\begin{aligned}\label{eqn_proof_thm_minimum_separation_A1}
    u_{p,q}(p, q) &= \sep^\alpha(\{p, q\}, \etal, \etau) = \sep^\alpha(X, \dl_X, \du_X).
\end{aligned}\end{equation}
Meanwhile, consider the $\alpha$-distance-reducing map constructed above and observe that $\ccalH$ satisfies the Axiom of Transformation (A2), and therefore for the given pair of distinct nodes $x, x' \in X$,
\begin{equation}\begin{aligned}\label{eqn_proof_thm_minimum_separation_A2}
    u_X(x, x') \geq u_{p, q} (\phixx(x), \phixx(x')) = u_{p, q}(p, q).
\end{aligned}\end{equation}
Since we can construct a $\alpha$-distance-reducing mapping $\phixx$ for any pair of nodes $x \neq x' \in X$, combining \eqref{eqn_proof_thm_minimum_separation_A1} and \eqref{eqn_proof_thm_minimum_separation_A2} yields
\begin{equation}\begin{aligned}\label{eqn_proof_thm_minimum_separation_final}
    u_X(x, x') \geq u_{p, q} (p, q) =  \sep^\alpha(X, \dl_X, \du_X), \forall x, x' \in X.
\end{aligned}\end{equation}
This is the definition of the Property of Minimum Separation (P1).

%
\section{Proof of Proposition \ref{prop_cluster_and_combine}} \label{apx_proof_prop_2}

Because $u_X^\CL$ is the output of single linkage to the symmetric dissimilarity $\alpha \cu_X(x, x') + (1 - \alpha) \cl_X(x, x')$, $u_X^\CL$ is a ultrametric. 

To see that axiom (A1) is fulfilled, pick an arbitrary two node space $\Delta_2(\dl, \du)$ and denote $(\{p, q\}, u_{p,q}^\CO) = \ccalH^\CO(\Delta_2(\dl, \du))$ as the output of applying combine-and-cluster clustering method to $\Delta_2(\dl, \du)$. It then follows that $\cu_{p,q}(p,q) = \du$ and $\cl_{p,q}(p,q) = \dl$. Also, because every possible chain from $p$ to $q$ must include a link from $p$ to $q$, the definition in \eqref{eqn_cluster_and_combine} becomes $u_{p,q}^\CL(p, q) = \alpha \du + (1 - \alpha) \dl$, which shows that axiom (A1) is satisfied.

To verify axiom (A2), consider arbitrary points $x, x' \in X$ and denote $C^\star(x, x')$ the chain achieving minimum cost in \eqref{eqn_cluster_and_combine_hhatc},
\begin{equation}\begin{aligned}\label{eqn_cluster_and_combine_A2_begin}
    u_X^\CL(x, x') = & \max_{i \mid x_i \in C^\star(x, x')} \hhatc_X(x_i, x_{i+1}),
\end{aligned}\end{equation}
Examine the transformed chain $C_Y(\phi(x), \phi(x'))$; since the map $\phi$ is $\alpha$-distance-reducing, it satisfies $\hhatc_Y(\phi(x_i), \phi(x_{i+1})) \leq \hhatc_X(x_i, x_{i+1})$ [cf. \eqref{eqn_distance_reducing_c}] for any link. Therefore, we can write
\begin{equation}\begin{aligned}\label{eqn_cluster_and_combine_A2_link}
    \max_{i \mid \phi(x_i) \in C_Y(\phi(x), \phi(x'))} &\hhatc_Y(\phi(x_i), \phi(x_{i+1})) \\
        & \quad \leq \max_{i \mid x_i \in C^\star(x, x')} \hhatc_X(x_i, x_{i+1}).
\end{aligned}\end{equation}
Further observe that $u_Y^\CL(\phi(x), \phi(x'))$ cannot exceed the cost in the given chain $C_Y(\phi(x), \phi(x'))$. Hence, 
\begin{equation}\begin{aligned}\label{eqn_cluster_and_combine_A2_final}
    u_Y^\CL(\phi(x), \phi(x')) \!\leq\! \max_{i \mid x_i \in C^\star(x, x')} \hhatc_X(x_i, x_{i+1}) \!=\! u_X^\CL(x, x'),
\end{aligned}\end{equation}
where the equality follows from \eqref{eqn_cluster_and_combine_A2_begin}. This shows $u_X^\CL$ satisfies axiom (A2) as in \eqref{eqn_axiom_of_transformation_ultrametric} and concludes the proof.

%

%

%

%

%

%
\urlstyle{same}
\bibliographystyle{IEEEtran}
\bibliography{clustering_biblio}

\end{document}